\documentclass[aps,showpacs]{revtex4}
\usepackage{epsf,epsfig,graphicx}
\usepackage{graphicx}
\usepackage{cancel,slashed}                             
\usepackage{amsmath}
\usepackage{mathtools}

\begin{document}
\title{Glauber gluons in pion-induced Drell-Yan processes revisited}

\author{Hsiang-nan Li} \email{hnli@phys.sinica.edu.tw}

\affiliation{Institute of Physics, Academia Sinica, Taipei,
Taiwan 115, Republic of China}

\begin{abstract}
We reanalyze the anomalous angular distribution of lepton pairs produced 
in a pion-induced Drell-Yan process by taking into account the Glauber 
gluon effect in the $k_T$ factorization theorem. Compared to the previous 
study, we adopt the realistic parton distribution functions (PDFs) for a proton 
from the CTEQ and for a pion from the xFitter, include the QCD evolutions of the 
strong coupling and the PDFs, and integrate the differential cross section over 
the kinematic regions for the NA10, E615 and COMPASS experiments. These 
improvements then allow rigorous confrontations of theoretical results with the data. 
It is shown that the lepton angular distribution and the violation of the Lam-Tung 
relation measured in all the above experiments can be well accommodated with a 
single Glauber phase. We illustrate the Glauber effect in the geometric picture 
for a Drell-Yan process, and its distinction from the conventional Boer-Mulders 
mechanism. The observables are pointed out, which can be used to discriminate the 
two proposals, when data become more precise.

\end{abstract}


\maketitle

The anomalous angular distribution of lepton pairs produced in a pion-induced Drell-Yan process
has been a long-standing puzzle. To explain this anomaly, we write the relevant differential 
cross section as \cite{RTO,LT78}
\begin{eqnarray}
\frac{1}{\sigma}\frac{d\sigma}{d\Omega}=\frac{3}{4\pi}\frac{1}{\lambda+3}
\left(1+\lambda \cos^2\theta+ \mu \sin 2\theta\cos\phi +
\frac{\nu}{2}\sin^2\theta \cos 2\phi\right),\label{diff}
\end{eqnarray}
with $d\Omega\equiv d\cos\theta d\phi$, where $\theta$ ($\phi$) is  the polar 
(azimuthal) angle of one of the leptons in the Collins-Soper (CS) frame \cite{CS77}.
We consider the process at an intermediate lepton-pair invariant mass $Q$, to which the
virtual photon contribution dominates over the $Z$ boson one. The coefficients $\lambda$ and 
$\nu$ are supposed to obey the Lam-Tung (LT) relation $\delta\equiv 2\nu+\lambda-1=0$ \cite{LT78} at a 
low lepton-pair transverse momentum $q_T$, that has been shown to hold largely 
under perturbative corrections \cite{Mirkes:1994dp,Qiu} and under 
parton-transverse-momentum and soft-gluon effects \cite{CK81,CB86}. Though the LT relation was
verified experimentally in the proton-proton and proton-deuteron Drell-Yan processes \cite{E866}, 
significant violation in the pion-induced ones was observed by the NA10 \cite{NA10} and
E615 \cite{E615}, and recently by the COMPASS \cite{COMPASS}: the substantial
deviation from $\delta=0$ clearly increases with $q_T$.

The above anomaly has stimulated  extensive theoretical investigations on its origin, which mainly 
resort to nonperturbative mechanisms \cite{BNM93,BNM05,BBK,EHVV,Boer99,BBH,LM04,BBNU05,Gamberg:2005ip,
Chang:2013pba,Zhou:2009rp}. For example, the vacuum effect proposed in \cite{BNM93,BNM05} causes the 
transverse-spin correlation between colliding partons, and the Boer-Mulders (BM) functions \cite{Boer99} 
introduce the spin-transverse-momentum correlation of a parton in an unpolarized hadron. As 
pointed out in \cite{BBNU05}, the vacuum effect is flavor-blind, so it is difficult to differentiate 
the pion-proton and proton-proton processes. The proposal based on the BM functions can differentiate 
these two processes, because a colliding anti-quark is a valence parton in a pion, but a sea parton in 
a proton \cite{LS10}. Note that the BM functions resolve the violation of the LT relation by 
increasing the coefficient $\nu$ in Eq.~(\ref{diff}) without changing $\lambda$. Our resolution 
\cite{Chang:2013pba} relies on infrared Glauber gluons appearing in the $k_T$ factorization theorem 
for complicated QCD processes \cite{CQ07,CQ06}, whose effect might be significant due to the unique 
role of a pion as a Nambu-Goldstone (NG) boson and a $q\bar q$ bound state simultaneously \cite{NS08}. 
The Glauber effect can modify the perturbative results of both $\lambda$ and $\nu$, and account for 
the LT violation observed in the pion-induced Drell-Yan process. An anti-proton is not a NG boson, so its
associated Glauber effect is expected to be weak, and the LT relation should be respected. It was thus 
suggested that examining the LT relation in a proton-anti-proton Drell-Yan process at low $q_T$ 
could discriminate the two mechanisms \cite{Chang:2013pba}: if violation is (not) observed, 
our (BM) proposal is irrelevant.

In this paper we will elaborate the proposal based on the Glauber gluon effect, and confront it
with the data, especially the preliminary COMPASS data \cite{COMPASS}, for the pion-induced Drell-Yan 
process. The purpose of our previous study \cite{Chang:2013pba} was to demonstrate the phenomenological 
impacts of the Glauber effect, and to estimate the LT violation at fixed rapidity $y$ and lepton-pair 
invariant mass $Q$. This is the reason why naive models for the parton distribution functions (PDFs) 
of a proton and a pion were employed in the factorization formulas to evaluate the angular 
coefficients. Hence, the theoretical results presented in \cite{Chang:2013pba} might not be compared 
with the data seriously. In the present work we will adopt the realistic PDFs for a proton from the 
CTEQ (CT18) \cite{Hou:2019efy} and for a pion from the xFitter \cite{Novikov:2020snp}. The latter are 
similar to those from the JAM \cite{Barry:2018ort}. We will also implement the QCD evolutions of the 
strong coupling $\alpha_s$ in the hard kernels involved in the factorization formulas and of the PDFs, 
and integrate the differential cross sections over the kinematic regions considered in 
different experiments. As observed in \cite{Chang:2018pvk,Chang:2019amo}, the theoretical outcomes 
for the three angular coefficients are sensitive to the variation of $Q$ actually. The above 
improvements then allow rigorous confrontations of our results for low $q_T$ spectra with the 
data. It will be shown that the Glauber effect enhances both $\lambda$ and $\nu$ in the perturbation 
theory \cite{Lambertsen:2016wgj,Boer:2006eq}, and leads to a better agreement with the NA10 data 
\cite{NA10}. The perturbative results for $\mu$ remain small under the Glauber effect, and match the 
data within experimental uncertainties. We then make predictions for the E615 \cite{E615} and COMPASS 
\cite{COMPASS} measurements with the same Glauber effect, and confirm that the observed LT violation is 
also accommodated.

\begin{figure}[tb]
\begin{center}
\includegraphics[scale=0.45]{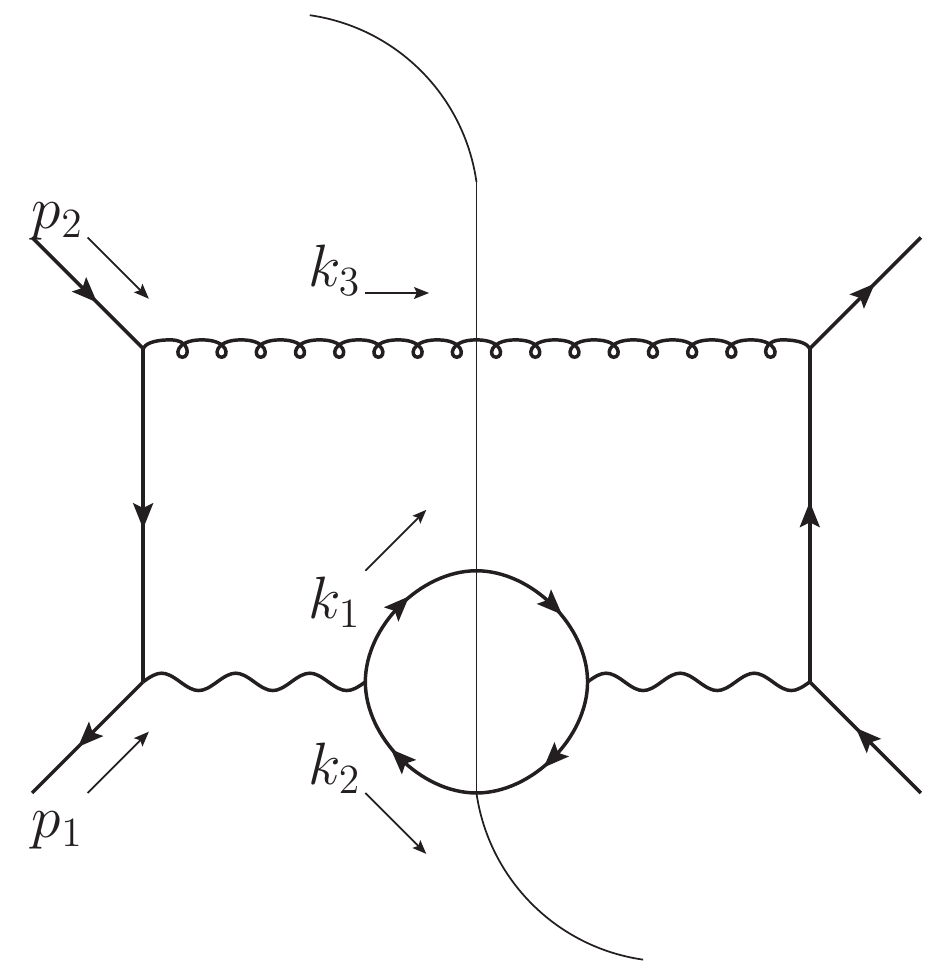}\hspace{0.5 cm}
\includegraphics[scale=0.45]{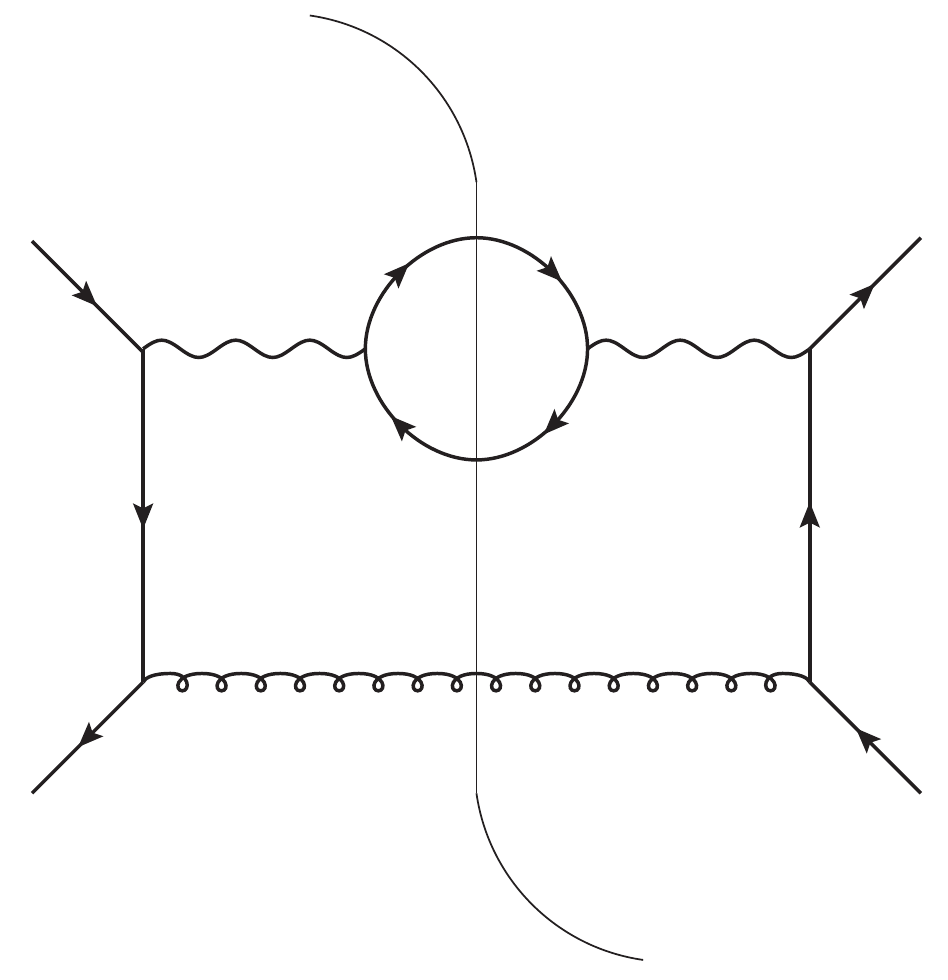}\hspace{0.5 cm}
\includegraphics[scale=0.45]{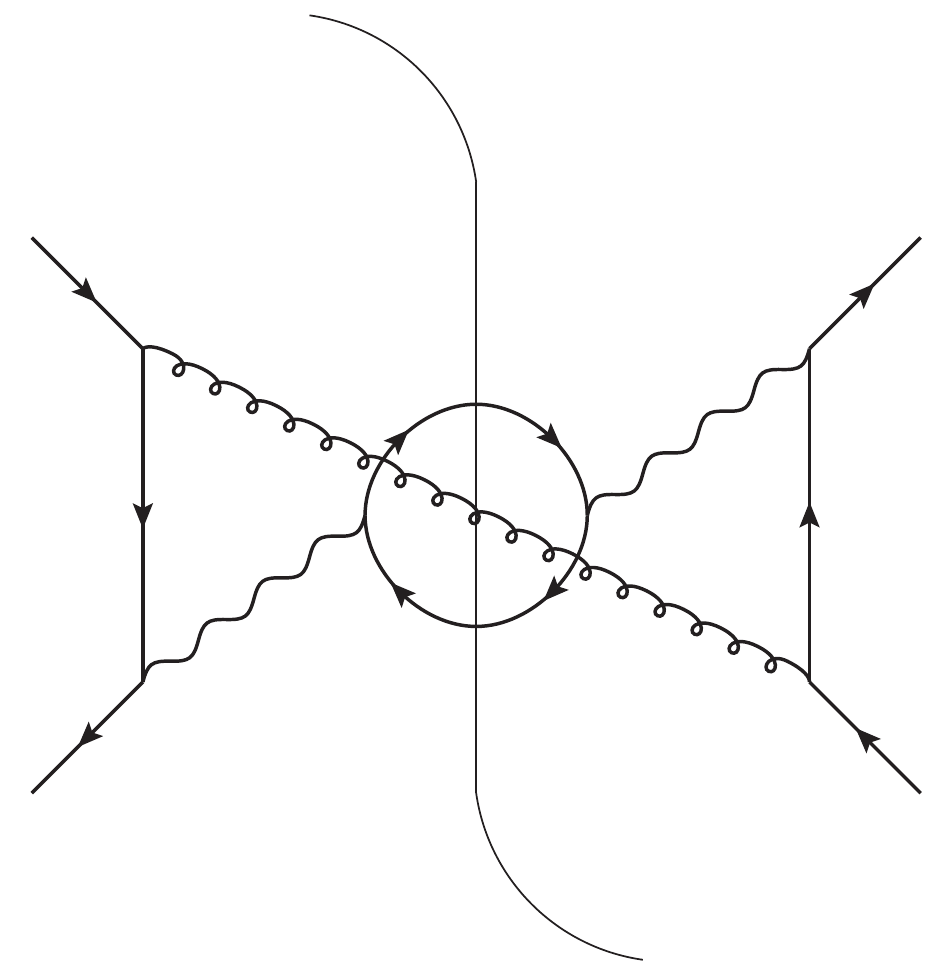}

(a) \hspace{4.3 cm} (b)\hspace{4.3 cm} (c)
\caption{LO diagrams for $\bar q(p_1)+q(p_2)\to\ell^-(k_1)+\ell^+(k_2)+g(k_3)$ in 
the pion-proton Drell-Yan process, where the variables in the parentheses 
label the parton momenta.}\label{fig1}
\end{center}
\end{figure}

It has been found \cite{Boer:2006eq} that the contribution to the aforementioned Drell-Yan processes
from the $q\bar q$ (quark-anti-quark) channel is more important than from the $qg$ (quark-gluon) 
channel. For instance, the former contributes more than 80\% of the total cross section for the 
COMPASS kinematics \cite{Chang:2018pvk}. This observation is reasonable, since the region with large 
parton momentum fractions dominates in fixed-target experiments, where gluonic partons have 
smaller distributions. Besides, currently available data are not precise enough for determining 
the sea and gluon distributions unambiguously \cite{Novikov:2020snp}. It has been verified that 
the coefficients $\lambda$ and $\nu$ are rather insensitive to resummation effects 
\cite{Lambertsen:2016wgj} and to next-to-leading-order (NLO) corrections at small $q_T\le 3$ GeV 
\cite{Lambertsen:2016wgj,Chang:2018pvk}, which we are focusing 
on. Therefore, we will confine ourselves to the leading-order (LO), i.e., $O(\alpha_s)$ $q\bar q$ 
contribution without the resummation in the investigation below. Note that tungsten was used 
in all the experiments involving pions \cite{NA10,E615,COMPASS}, but we will not take into account 
nuclear effects as in \cite{Lambertsen:2016wgj}. The LO parton-level diagrams for the scattering 
$\bar q(p_1)+q(p_2)\to\ell^-(k_1)+\ell^+(k_2)+g(k_3)$ in the pion-proton Drell-Yan process, where 
the variables in the parentheses label the parton momenta with $k_3 = p_1 + p_2 - k_1 - k_2$, are 
displayed in Fig.~\ref{fig1}. The momentum $p_1$ ($p_2$) with the dominant plus (minus) component 
is carried by the valence anti-quark (quark) in the pion (proton). The explicit expressions of 
the corresponding hard kernels are referred to \cite{KLR78,LT79,JC79,CK79,JL79,CB86}.


We first briefly review the appearance of infrared Glauber divergences in radiative corrections 
to Fig.~\ref{fig1}. It is obvious that the low $q_T$ spectra of lepton pair productions in a 
Drell-Yan process meet the necessary conditions for the existence of Glauber gluons: the 
$k_T$ factorization theorem is the appropriate theoretical framework for the low $q_T$ 
spectra, in which the dependence on parton transverse momenta should be kept; a 
final-state parton is required to balance the lepton-pair $q_T$, so at least 
three partons participate the hard scattering; the lepton-pair momentum $q=k_1+k_2$ is restricted 
in a finite phase space, such that the final-state parton is not fully inclusive in kinematics, 
and the Glauber divergences in various diagrams do not cancel exactly. We stress that a final-state
parton is needed to help balance $q_T$, as $q_T$ is about few GeV, the region where
the LT violation is significant. The intrinsic transverse momenta of the initial-state partons alone
are insufficient to sustain such $q_T$. According 
to the elucidation in \cite{CQ07,CQ06,CL09,Chang:2013pba}, radiative gluons emitted by a 
spectator line in the pion (like a rung gluon that can be exchanged between the two anti-quarks of
the momentum $p_1$ in Fig.~\ref{fig1}), and attaching to lines in other 
subprocesses produce Glauber divergences. For the diagrams in Fig.~\ref{fig1}, the Glauber 
divergences are extracted from the attachments to the quark of the momentum $p_2$ in the 
proton, to the gluon of the momentum $k_3$, and to the vertical quark lines \cite{CL09}. 
Note that a Glauber gluon, different from an ordinary soft gluon, gives rise to an imaginary 
infrared logarithm. To get a real cross section, at least two Glauber gluons are present, which 
may be located on the same side or on the opposite sides of the final-state cut. It has been 
shown that the infrared divergences from these two types of gluon allocations do not cancel 
exactly in the $k_T$ factorization theorem \cite{CQ06}, and that the imaginary Glauber 
logarithms can be factorized into a universal nonperturbative phase factor to all orders in 
$\alpha_s$ at low $q_T$ \cite{CL09}.

The transverse momentum $l_T$ of a Glauber gluon flows through the parton-level hard kernels
represented by the two vertical quark lines in Fig.~\ref{fig1} \cite{CL09}. The two vertical 
quarks in Fig.~\ref{fig1}(a) have small invariant masses in the positive rapidity region of 
the lepton pair, and those in Fig.~\ref{fig1}(b) have small invariant masses in the 
negative rapidity region. However, the two vertical quarks in Fig.~\ref{fig1}(c) cannot have
small invariant masses simultaneously: the quark on the left-hand (right-hand) side of the 
final-state cut has a small (large) invariant mass, as the lepton pair is produced with 
positive rapidity. This difference in the hard kernels, as the differential cross section is 
integrated over the rapidity, renders the net Glauber effect from the two sides of the 
final-state cut suppressed for Fig.~\ref{fig1}(c) compared to those for 
Figs.~\ref{fig1}(a) and \ref{fig1}(b). To elaborate the above statement, we quote the 
factorization formula in the impact-parameter space for a Drell-Yan process with Glauber 
gluon exchanges \cite{CL09}
\begin{eqnarray}
& &\int d^2b_ld^2b_rd^2b'_ld^2b'_r
e^{-iS({\bf b}_l)}H({\bf b}_l-{\bf b}'_l,{\bf
b}_r-{\bf b}'_r)e^{iS({\bf b}_r)}\Phi_{\pi}({\bf b}'_l,{\bf b}'_r)
e^{-i{\bf q}_{T}\cdot({\bf b}_l-{\bf b}_r-{\bf b}'_l+{\bf
b}'_r)}\cdots,\label{sig1}
\end{eqnarray}
where only the relevant factors are shown explicitly, and the exponentials $e^{\pm iS}$ 
organize the Glauber gluons to all orders in $\alpha_s$. In the presence of the Glauber 
gluons that carry transverse momenta, both the Fourier-transformed hard kernel $H$ 
and transverse-momentum-dependent (TMD) pion PDF $\Phi_\pi$ depend on two impact parameters. 
That is, the partons on the left-hand (labelled by the subscripts $l$) and right-hand 
(labelled by the subscripts $r$) sides of the final-state cut have different transverse coordinates.

It is easy to see from Eq.~(\ref{sig1}) that both the arguments ${\bf b}_l-{\bf b}'_l$ and 
${\bf b}_r-{\bf b}'_r$ can be large, when the two vertical quarks have small invariant masses
as they do in Figs.~\ref{fig1}(a) and \ref{fig1}(b). Namely, ${\bf b}_l$ (${\bf b}_r$) is 
different from ${\bf b}'_l$ (${\bf b}'_r$), and takes a value in a wide range, so there is no strong 
cancellation between the Glauber factors $e^{-iS({\bf b}_l)}$ and $e^{iS({\bf b}_r)}$ from the 
two sides of the final-state cut. When one of the vertical quarks, say, the one on the right-hand 
side of the final-state cut has a large invariant mass, the region with small ${\bf b}_r-{\bf b}'_r$, 
i.e., with ${\bf b}_r\approx{\bf b}'_r$ dominates. For a finite $q_T$ of order 1 
GeV, the Fourier factor in Eq.~(\ref{sig1}), $\exp[-i{\bf q}_{T}\cdot({\bf b}_l-{\bf b}_r-{\bf b}'_l+{\bf
b}'_r)]\approx \exp[-i{\bf q}_{T}\cdot({\bf b}_l-{\bf b}'_l)]$, enforces the condition that ${\bf b}_l$ 
cannot be very different from ${\bf b}'_l$. It turns out that both ${\bf b}_l$ and  ${\bf b}_r$
are restricted in the support of ${\bf b}'_l$ and ${\bf b}'_r$ for $\Phi_{\pi}({\bf b}'_l,{\bf b}'_r)$
defined by the transverse extent of the pion. The cancellation between $e^{-iS({\bf b}_l)}$ and 
$e^{iS({\bf b}_r)}$ then becomes stronger, explaining why the net Glauber effect is minor for 
Fig.~\ref{fig1}(c). Below we will neglect the Glauber effect on Fig.~\ref{fig1}(c), and assume that 
Figs.~\ref{fig1}(a) and \ref{fig1}(b) acquire an additional factor $\cos S$ \cite{Chang:2013pba}. 
The Glauber phase $S$ is proportional to the product of $\alpha_s$ and an infrared logarithm, 
if computed in the perturbation theory. The expansion of the Glauber factor 
$\cos S$ in powers of $\alpha_s$ reflects the fact that an odd number of Glauber gluons does not 
contribute to a real cross section. Because the Glauber phase is of nonperturbative origin, and its 
explicit expression is unknown, we simply treat $S$ as a constant, which parametrizes the Glauber 
effect averaged over the impact parameters, i.e., over the internal transverse momenta. The complexity of the 
analysis is thus greatly reduced by avoiding the lengthy convolution in Eq.~(\ref{sig1}).  
The simplified factorization formulas with the average Glauber phase $S$ are derived in detail
in the Appendix.

It has been pointed out \cite{Liu:2015sra} that a Glauber factor, despite being universal 
once the $k_T$ factorization is established, generates different effects in different processes. 
The reason is that a Glauber factor makes its impact through the convolution with other 
subprocesses, including TMD hadron wave functions. As 
demonstrated in \cite{Liu:2015sra}, the pion ($\rho$ meson) TMD wave function with a weak (strong) 
falloff in a parton transverse momentum leads to significant (moderate) Glauber effects on 
two-body hadronic $B$ meson decays. This observation is consistent with the dual role of a pion 
as a massless NG boson and as a $q\bar q$ bound state, which requires a tighter spatial 
distribution for its leading Fock state. The Glauber effect 
has been introduced to resolve several puzzling data in two-body hadronic heavy flavor decays 
into pions, such as the abnormally large $B^0\to\pi^0\pi^0$ and $\pi^0\rho^0$ branching ratios 
\cite{LM06,LM11,Li:2014haa}, the very different direct $CP$ asymmetries in the $B^+\to\pi^0 K^+$ 
and $B^0\to\pi^- K^+$ decays \cite{LMS05,Liu:2015upa}, and the difference between the 
$D^0\to\pi^+\pi^-$ and $K^+K^+$ branching ratios that exceeds the expected SU(3) symmetry 
breaking \cite{diag,LLY12,Li:2021req}. It has been elaborated recently that the data of the 
$D\to \pi\pi$ and $\pi K$ branching ratios reveal prominent Glauber effects \cite{Li:2021req}.


We start with the differential cross section for the pion-proton Drell-Yan process
\begin{eqnarray}
\frac{d\sigma}{dQ^2dy dq_T^2d\Omega}&=&\frac{N}{s^2}
\bigg[H_0(Q^2,y,q_T^2)+H_\lambda(Q^2,y,q_T^2) \cos^2\theta
+H_\mu(Q^2,y,q_T^2) \sin 2\theta\cos\phi\nonumber\\
& &\hspace{0.6cm}+\frac{1}{2}H_\nu(Q^2,y,q_T^2) \sin^2\theta \cos 2\phi\bigg],
\label{hi}
\end{eqnarray}
where the normalization constant $N$ is irrelevant to the evaluations of the angular
coefficients, and $s$ is the center-of-mass energy squared. The functions $H_i$, 
$i=0$, $\lambda$, $\mu$ and $\nu$, are written as the convolutions of the hard kernels 
$\hat H_i$ with the pion PDF $\phi_\pi$ and the proton PDF $\phi_P$ at the scale $\mu=Q$,
\begin{eqnarray}
H_i(Q^2,y,q_T^2)&=& \frac{\alpha_s(Q^2)}{Q^2}\int dx_1dx_2 \phi_\pi(x_1,Q^2)
\hat H_i(x_1,x_2,Q^2,y,q_T^2)\phi_P(x_2,Q^2)\nonumber\\
& &\hspace{1.5cm}\times \delta\left(x_1x_2-(x_1e^{-y}+x_2e^y)
\frac{\sqrt{Q^2+q_T^2}}{\sqrt{s}}+\frac{Q^2}{s}\right).\label{4}
\end{eqnarray}
The $\delta$ function, arising from the on-shell condition $k_3^2=0$, specifies the 
relation between the parton momentum fractions $x_1$ and $x_2$. It has been checked that 
the alternative choice $\mu=\sqrt{Q^2+q_T^2}$ yields very similar results even at LO 
\cite{Lambertsen:2016wgj}. The angular coefficients in Eq.~(\ref{diff}) and the LT violation $\delta$ 
are defined by
\begin{eqnarray}
\lambda,\mu,\nu,\delta=\frac{\int dQ^2dy H_{\lambda,\mu,\nu,\delta}(Q^2,y,q_T^2)}
{\int dQ^2dy H_0(Q^2,y,q_T^2)},
\label{ang}
\end{eqnarray}
where the factorization formula for $H_\delta$ is similar to Eq.~(\ref{4}) with the hard kernel 
$\hat H_\delta$. We point out that $\nu$ is more sensitive to the changes of PDFs than $\lambda$, and
that $\mu$ is equal to zero, when the pion and proton PDFs have the same functional 
form \cite{Chang:2013pba}. 

To present the expressions of the LO hard kernels $\hat H_i$, we first choose the parton and 
lepton momenta in the CS frame as
\begin{eqnarray}
& &p_1 = E_1( 1, -\sin\theta_1, 0, \cos\theta_1), \hspace{1.6cm}
p_2 = E_2( 1, -\sin\theta_1, 0, -\cos\theta_1 ), \nonumber\\
& &k_1 = k(1, \sin\theta\cos\phi, \sin\theta\sin\phi, \cos\theta),\;\;\;\;
k_2 = k(1, -\sin\theta\cos\phi, -\sin\theta\sin\phi, -\cos\theta),\label{pp}
\end{eqnarray}
where $E_1$ and $E_2$ are the parton energies, $k$ is the lepton energy, and $\theta_1$ is the 
angle between the momentum $\bf p_1$ and the $z$ axis. In terms of the kinematic variables in 
Eq.~(\ref{pp}), $\hat H_i$ from the $q\bar q$ channel modified by the Glauber factor $\cos S$ read 
\begin{eqnarray}
\hat H_0&=&\left(\frac{E_1}{E_2}+\frac{E_2}{E_1}\right)
\left(\frac{1}{\sin^2\theta_1}+\frac{1}{2}\right)\nonumber\\
& &+(\cos{S}-1)\left[\left(\frac{2E_1E_2}{k^2}\cos^2\theta_1
-\frac{E_1}{E_2}-\frac{E_2}{E_1}\right)\left(\frac{1}{\sin^2\theta_1}-\frac{1}{2}\right)
+\left(\frac{k}{E_1}+\frac{k}{E_2}-2\right)\frac{2}{\sin^2\theta_1} \right]\\
%
\hat H_\lambda&=&
\left(\frac{E_1}{E_2}+\frac{E_2}{E_1}\right)\left(\cot^2\theta_1-\frac{1}{2}\right)\nonumber\\
& &+ (\cos{S}-1)\left[\left(\frac{E_1}{E_2}
+\frac{E_2}{E_1}-2\right)\left(\cot^2\theta_1-\frac{1}{2}\right)+\frac{E_1E_2}{k^2}\cos^2\theta_1
-1\right],
\\
\hat H_\mu &=&\left(\frac{E_2}{E_1}-\frac{E_1}{E_2}\right)\cot\theta_1 
+(\cos{S}-1)\left(\frac{E_1-E_2}{k}+\frac{E_2}{E_1}-\frac{E_1}{E_2}\right)\cot\theta_1,\\
\hat H_\nu&= &\left(\frac{E_1}{E_2}
+\frac{E_2}{E_1}\right)-(\cos{S}-1)\left(\frac{2E_1E_2}{k^2}\cos^2\theta_1-\frac{E_1}{E_2}
-\frac{E_2}{E_1}\right),\label{h}\\
\hat H_\delta&= &\frac{2(\cos S-1)}{\sin^2\theta_1}\left[\frac{E_1-k}{E_2}+\frac{E_2-k}{E_1}-
\left(\frac{E_1E_2}{k^2}\cos^2\theta_1-1\right)(1+\sin^2\theta_1)\right],
\end{eqnarray}
where those pieces multiplied by $\cos S-1$ arise from Figs.~\ref{fig1}(a) and \ref{fig1}(b).
It is seen that the hard kernel $\hat H_\delta$ for LT violation $\delta$ vanishes as $S=0$.
Compared to \cite{Chang:2013pba}, an overall factor $1/\sin^2\theta_1$, that depends on the lepton-pair 
invariant mass $Q$, has been included. This factor was neglected before, since it cancels in the ratios 
for defining the angular coefficients at fixed $Q$. Here we will integrate the differential cross 
section over kinematic variables in order to confront our results with the data rigorously.

We then transform the kinematic variables $E_1$, $E_2$, $k$, and $\theta_1$ in the CS
frame to those in the center-of-mass frame of the colliding hadrons via \cite{Chang:2013pba}
\begin{eqnarray}
k=\frac{Q}{2},\;\;\;
\sin\theta_1=\frac{q_T}{\sqrt{Q^2+q_T^2}},\;\;\;
E_1=\frac{e^{-y}}{\cos\theta_1}x_1P_1^0,\;\;\; E_2=\frac{e^y}{\cos\theta_1} x_2P_2^0,
\label{tra}
\end{eqnarray}
with the pion and proton energies $P_1^0=P_2^0=\sqrt{s}/2$, and obtain the hard 
kernels $\hat H_i(x_1,x_2,Q^2,y,q_T^2)$. It is found that $\sin\theta_1$ is proportional to 
the lepton-pair transverse momentum $q_T$, i.e., to the boost of the CS frame relative to the 
center-of-mass frame. The constraint on the gluon energy $k_3^0>0$ together with the on-shell 
condition $k_3^2=0$ favors the region of large momentum fractions,
\begin{eqnarray}
& &\frac{e^y\sqrt{Q^2+q_T^2}-Q^2/\sqrt{s}}{\sqrt{s}-e^{-y}\sqrt{Q^2+q_T^2}}
\le x_1 \le 1,
\end{eqnarray}
for an intermediate $Q$, which dominates in fixed-target experiments.

\begin{figure}[tb]
\begin{center}
\includegraphics[scale=0.5]{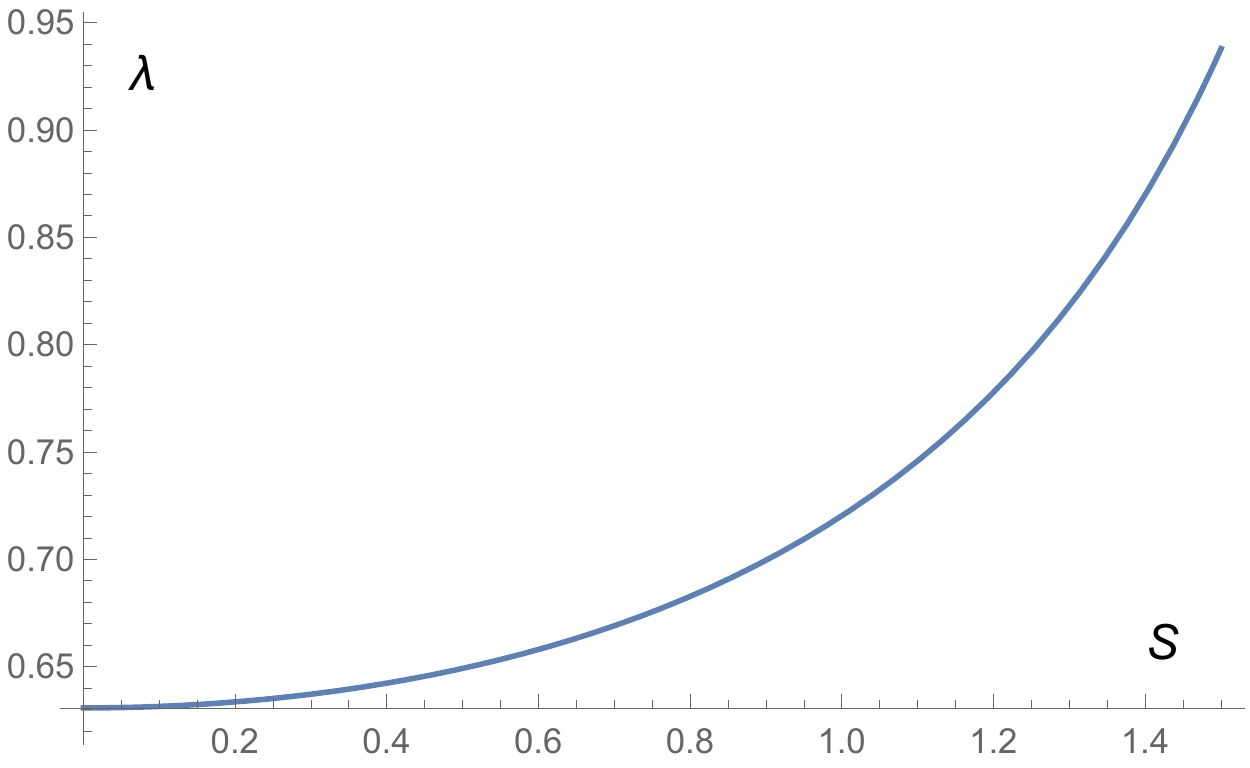}\hspace{0.5 cm}
\includegraphics[scale=0.5]{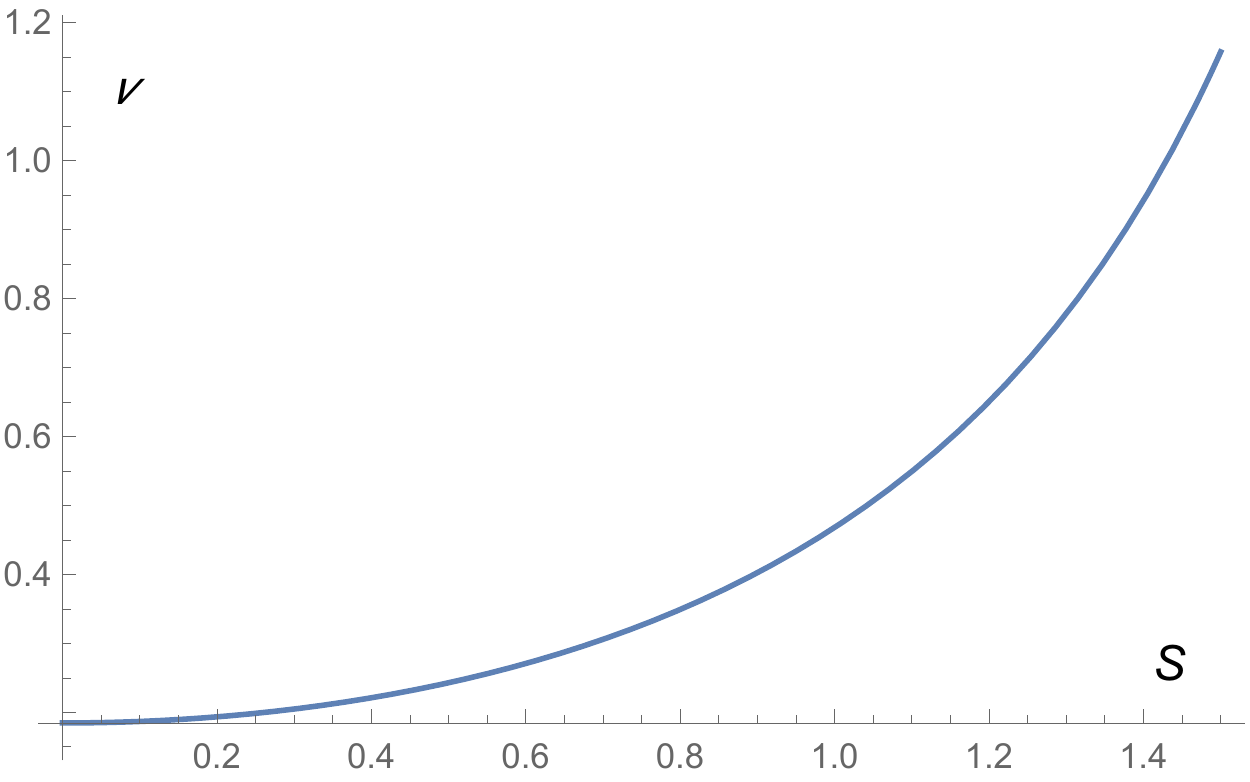}

\mbox{}\hspace{0.7 cm}(a) \hspace{6.3 cm} (b)

\caption{Dependencies of the angular coefficients (a) $\lambda$ and (b) $\nu$ on the
Glauber phase $S$ for the pion beam energy $E_\pi=194$ GeV 
and the lepton-pair transverse momentum $q_T=2.5$ GeV with the cuts 
in Eq.~(\ref{cut1}).}\label{fig2}
\end{center}
\end{figure}

We calculate the angular coefficients to be compared with the NA10 data \cite{NA10},
adopting the PDFs for a proton from the CT18 \cite{Hou:2019efy} and for a pion from the 
xFitter \cite{Novikov:2020snp}. The integrations in Eq.~(\ref{ang}) are performed over the 
range $Q\ge 4$ GeV for the pion beam energy $E_\pi=286$ GeV, over $Q\ge 4.05$ GeV for  
$E_\pi=194$ GeV, and over $Q\ge 4$ GeV for $E_\pi=140$ GeV with the bottomonium region 
8.5 GeV $\le Q\le 11$ GeV being excluded \cite{Lambertsen:2016wgj}. The cut $0\le x_\pi\le 0.7$ 
is also implemented with
\begin{eqnarray}
x_\pi=\frac{1}{2}\left(x_F+\sqrt{x_F^2+\frac{4Q^2}{s}}\right),
\end{eqnarray}
$x_F$ being the Feynman variable. The variable $x_\pi$ corresponds to the parton momentum 
fraction $x_1$, and $x_F$ ($\sqrt{x_F^2+4Q^2/s}$) is proportional to the longitudinal 
momentum (energy) of the lepton pair in the limit $k_3\to 0$ in the center-of-mass frame
of the colliding hadrons. The physical ranges of $Q$ and $y$ for a given $q_T$ are those, 
in which $x_1$ and $x_2$ take values between 0 and 1. The combination of the above 
kinematic constraints leads to the ranges
\begin{eqnarray}
& &\frac{1}{2}\left(b-\sqrt{b^2-4}\right)\le e^y \le
\frac{1}{2}\left(a+\sqrt{a^2+4}\right),\;\;\;\;Q^2\le 0.7 s\sqrt{1-\frac{4q_T^2}{(1-0.7^2)s}}\nonumber\\
& &\frac{1}{2}\left(b-\sqrt{b^2-4}\right)\le e^y \le
\frac{1}{2}\left(b+\sqrt{b^2-4}\right),\;\;\;\;0.7 s\sqrt{1-\frac{4q_T^2}{(1-0.7^2)s}}\le Q^2\le
s-2\sqrt{s}q_T,\nonumber\\
& &a=\frac{0.7^2 s-Q^2}{0.7\sqrt{s(Q^2+q_T^2)}},\;\;\;\;
b=\sqrt{\frac{s}{Q^2+q_T^2}}\left(1+\frac{Q^2}{s}\right).\label{cut1}
\end{eqnarray}
Equation~(\ref{cut1}) implies that the allowed range of $y$ shrinks with $Q^2$, and $y\to 0$
as $Q^2$ approaches to its upper bound  $s-2\sqrt{s}q_T$.

The dependencies of the angular coefficients $\lambda$ and $\nu$ on the Glauber phase $S$
for the pion beam energy $E_\pi=194$ GeV and the lepton-pair transverse momentum 
$q_T=2.5$ GeV under the cuts in Eq.~(\ref{cut1}) are displayed in Fig.~\ref{fig2}. 
It is found that the values of $\lambda$ and $\nu$ at $S=0$, i.e., the perturbative results 
without the Glauber effect, reproduce those in \cite{Lambertsen:2016wgj,Chang:2018pvk}. 
Namely, the simplification made in our calculation, i.e., considering only the LO $q\bar q$ 
channel is justified. It is interesting to see that the Glauber effect enhances both 
$\lambda$ and $\nu$, and the deviation from the LT relation $2\nu+\lambda-1=0$ is then
induced. As emphasized before, this feature differentiates our resolution to the LT violation
from the one based on the BM functions, which increases only $\nu$. Therefore, 
separate comparisons of theoretical predictions with future precise data of $\lambda$ and 
$\nu$ is likely to discriminate the two proposals. We observe that our results of $\nu$ are 
more sensitive to the variation of the Glauber phase than those of $\lambda$, and that the 
NA10 data for $\nu$ are more precise than for $\lambda$ (and also more precise than the E615 
and COMPASS data). We thus fix $S=0.8$ by collating Fig.~\ref{fig2}(b) and the NA10 data 
for $\nu$ at $E_\pi=194$ GeV and $q_T=2.5$ GeV, and employ this single input to make 
predictions for all other quantities. With $S=0.8$, the perturbative values of $\lambda$ 
are enhanced by 10\%, which is not as strong as obtained in our previous naive 
estimate \cite{Chang:2013pba}, and those of $\nu$ are enhanced by a factor of 2. 
We simply vary the Glauber phase to $S=0.7$ and $S=0.9$ to assess the theoretical uncertainties, 
which are about 5\% for $\lambda$ and 15\% for $\nu$. It is noticed that the angular coefficient 
$\mu$ remains tiny \cite{Chang:2013pba}: it takes the value $\mu=-0.028$ for $S=0.8$ and $q_T=2.5$ GeV, 
which is consistent with the NA10 data, and much smaller than the experimental errors. We will 
not present the results of $\mu$ hereafter.


\begin{figure}[tb]
\begin{center}
\includegraphics[scale=0.42]{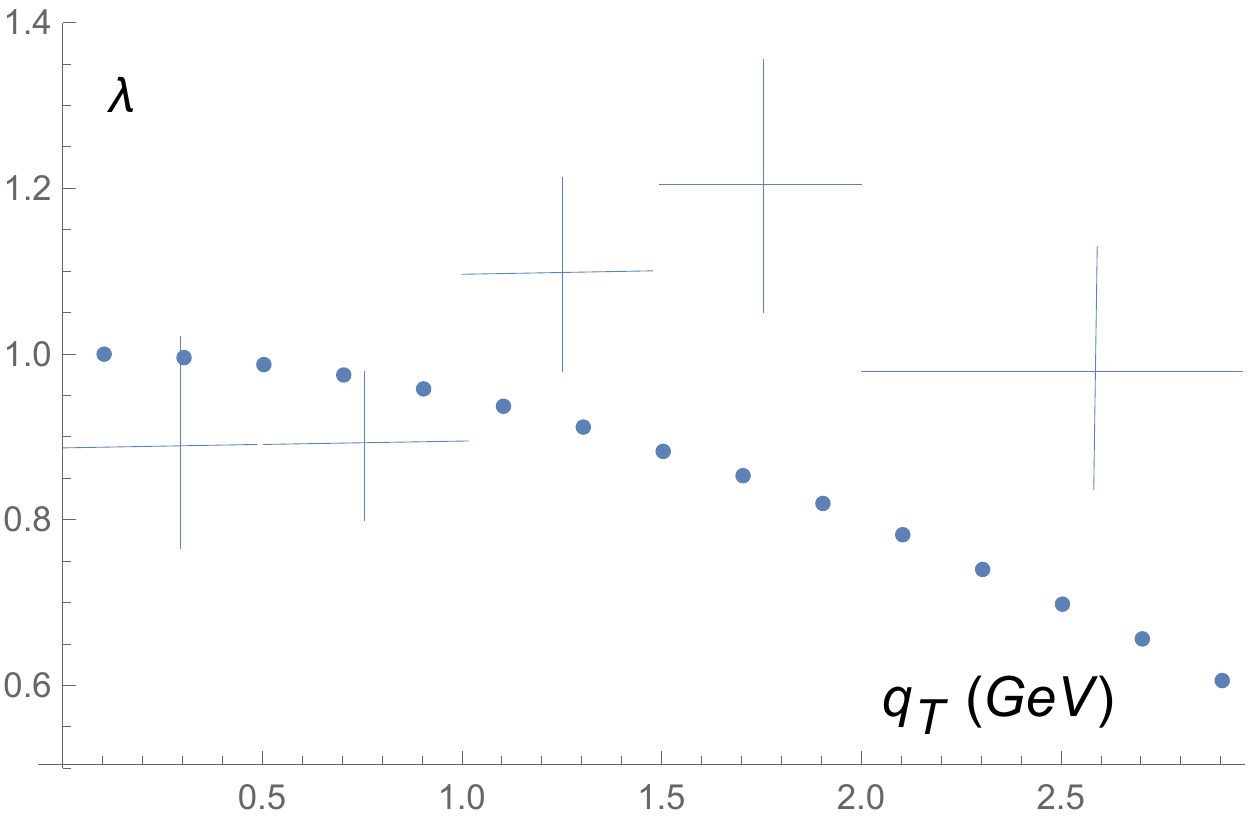}\hspace{0.3 cm}
\includegraphics[scale=0.42]{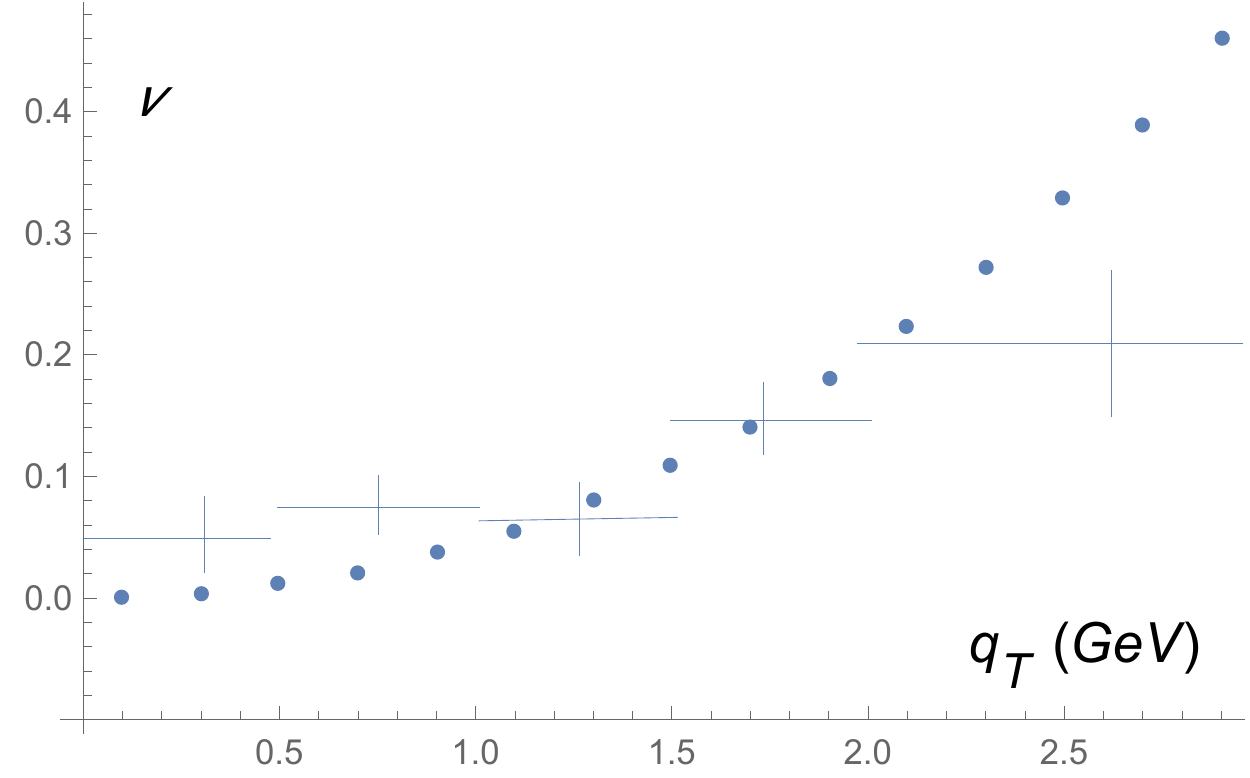}\hspace{0.3 cm}
\includegraphics[scale=0.42]{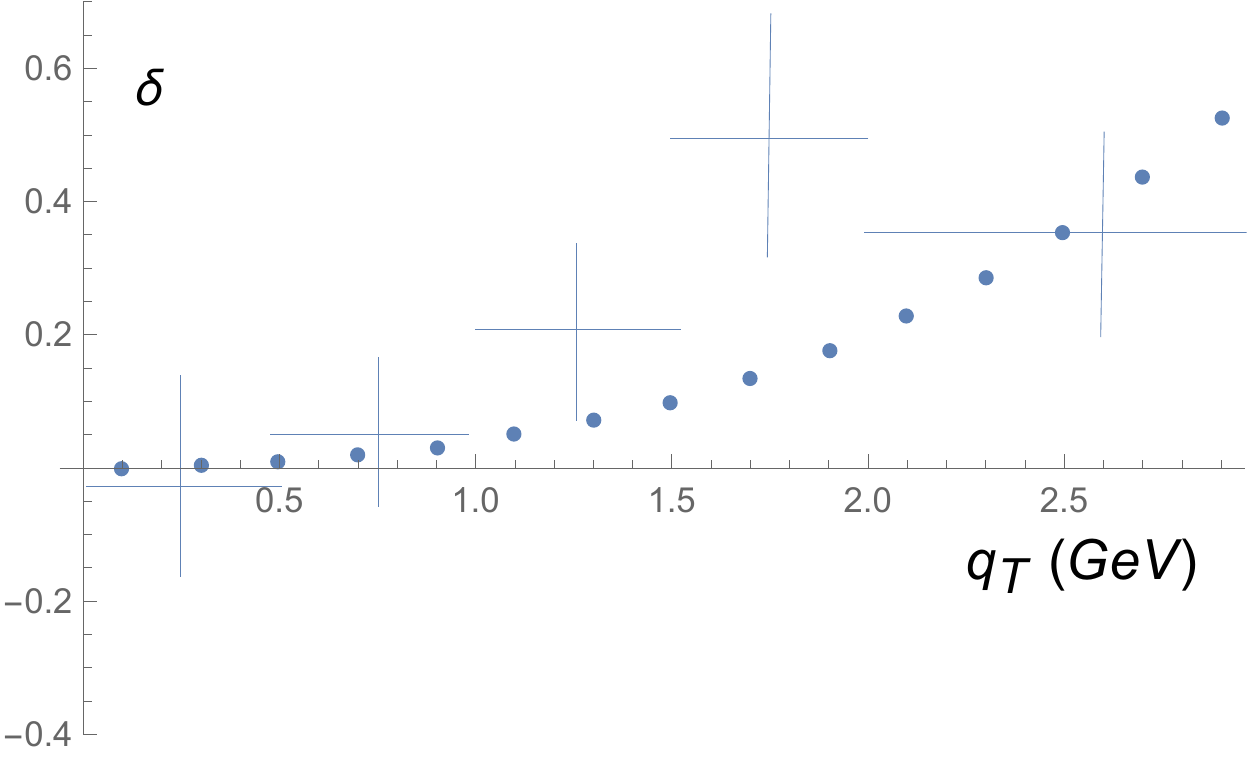}

(a) 

\includegraphics[scale=0.42]{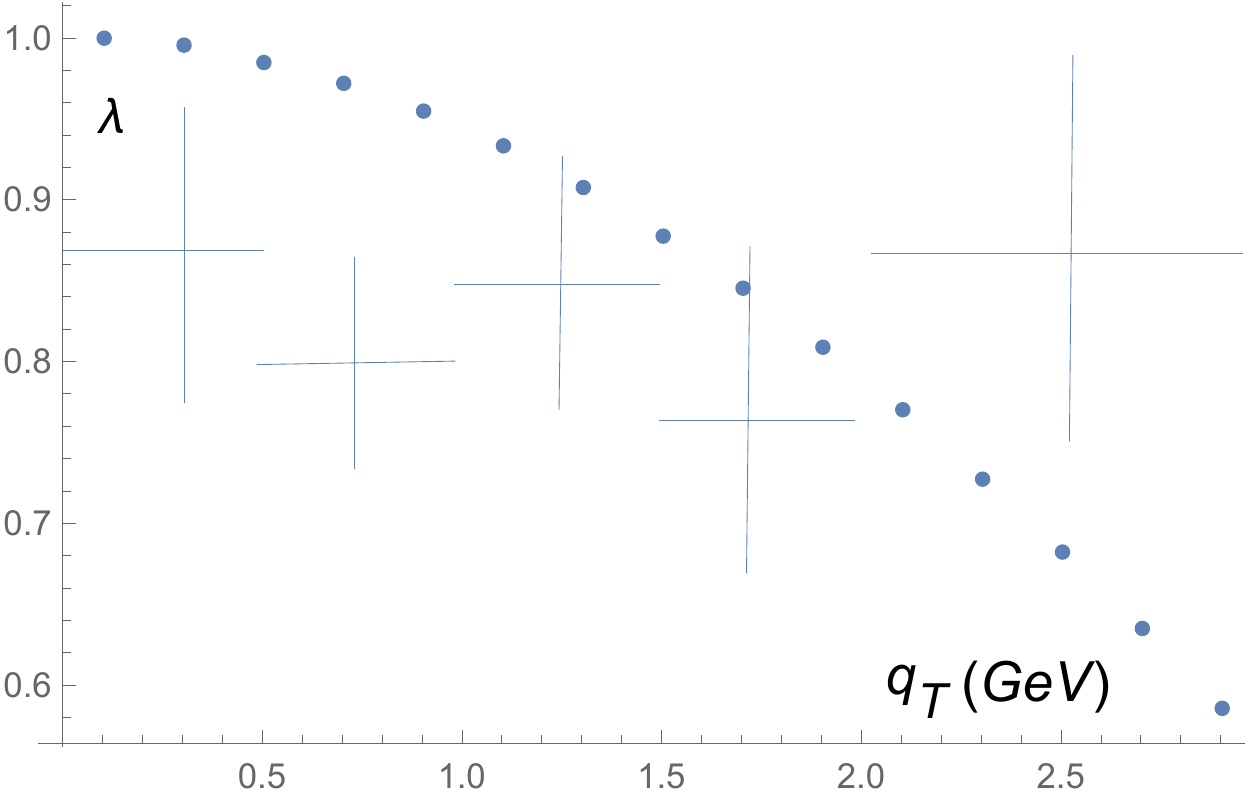}\hspace{0.3 cm}
\includegraphics[scale=0.42]{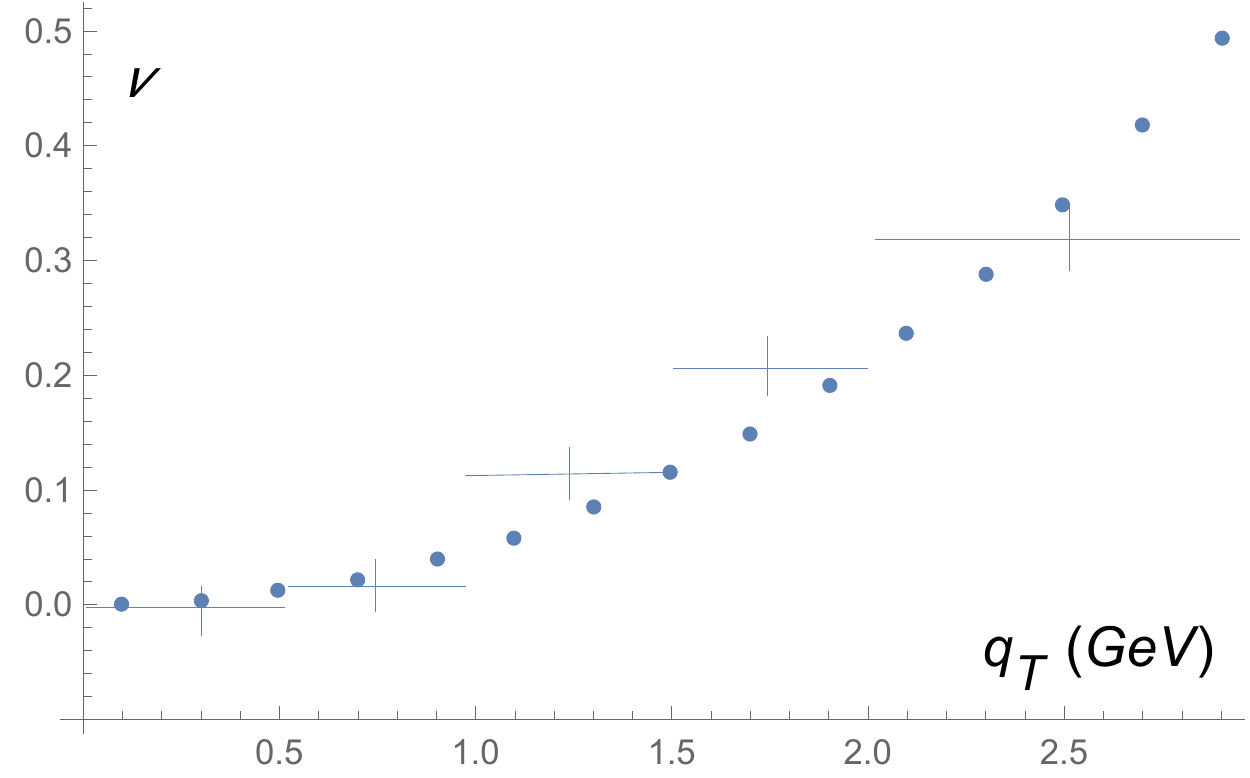}\hspace{0.3 cm}
\includegraphics[scale=0.42]{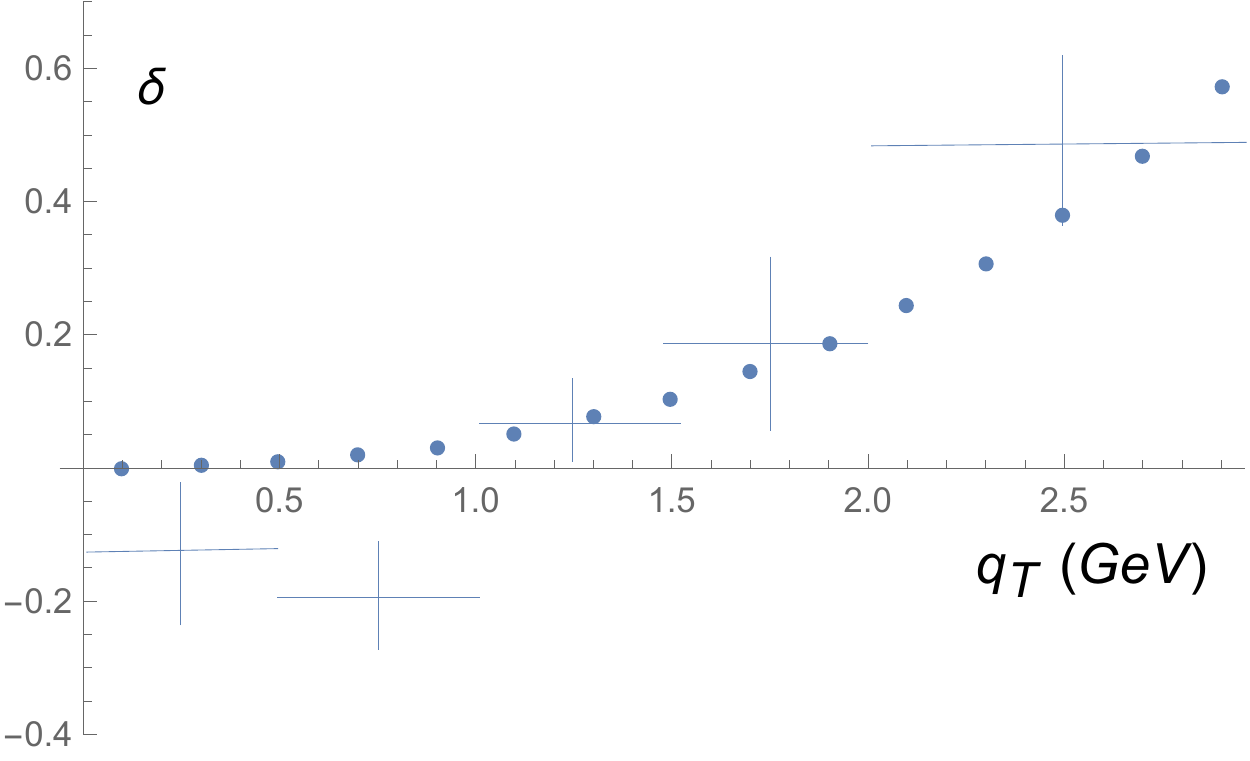}

(b) 

\includegraphics[scale=0.42]{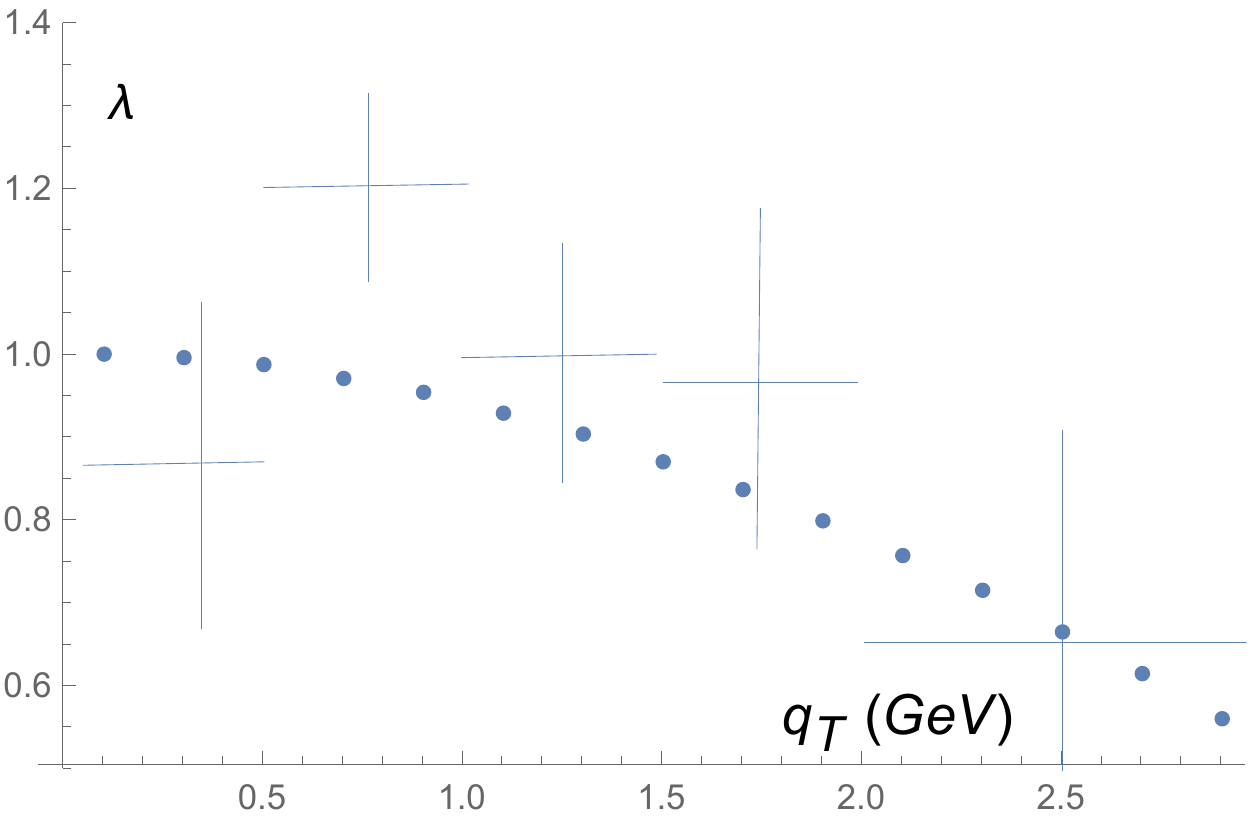}\hspace{0.3 cm}
\includegraphics[scale=0.42]{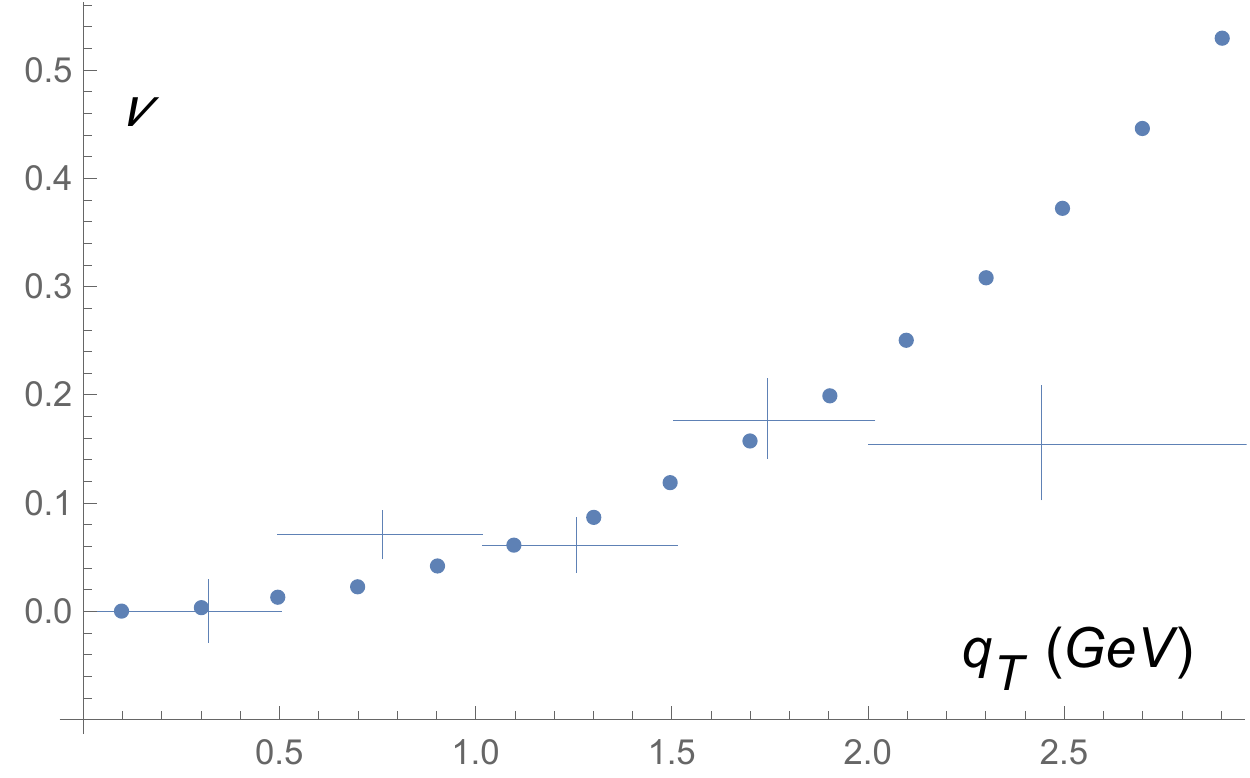}\hspace{0.3 cm}
\includegraphics[scale=0.42]{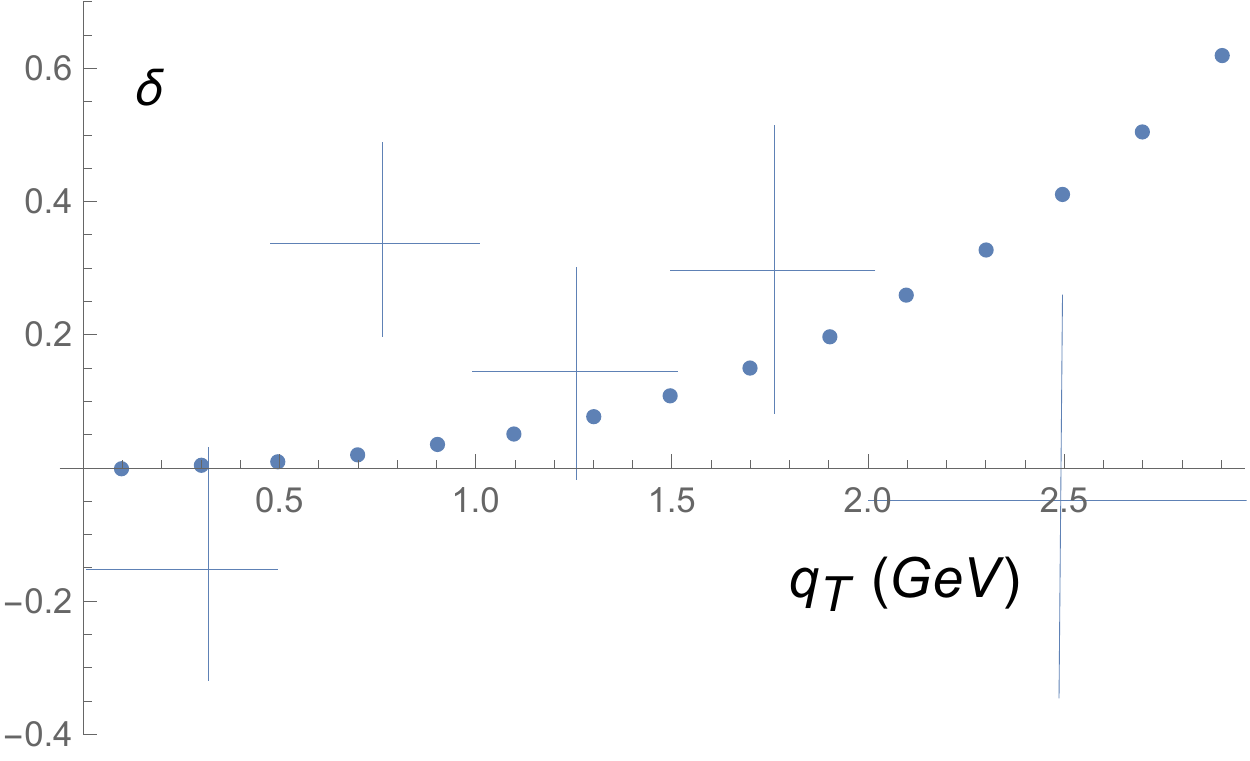}

(c) 

\caption{Dependencies of $\lambda$, $\nu$, and the LT violation 
$\delta\equiv 2\nu+\lambda-1$ on the lepton-pair transverse momentum $q_T$, and their comparisons 
with the NA10 data \cite{NA10} for the pion beam energies (a) $E_\pi=286$ GeV,
(b) $E_\pi=194$ GeV, and (c) $E_\pi=140$ GeV.}\label{fig3}
\end{center}
\end{figure}

The changes of the angular coefficients $\lambda$ and $\nu$, and the violation 
$\delta\equiv 2\nu+\lambda-1$ of the LT relation with the lepton-pair transverse momentum $q_T$ 
for the Glauber phase $S=0.8$ under Eq.~(\ref{cut1}) are exhibited in Fig.~\ref{fig3}. We focus on 
the low $q_T\le 3$ GeV region, for which the $k_T$ factorization theorem is more appropriate, and the 
Glauber effect is expected to be significant. Note that the curve of $\nu$ in Fig.~\ref{fig3}(b) will 
go below the data for $E_\pi=194$ GeV, if $S$ is set to 0.7, and those will go above the data for 
$E_\pi=286$ GeV and 140 GeV, if $S$ is set to 0.9. This check supports our choice $S=0.8$, which 
improves the overall agreement with the NA10 data \cite{NA10} of $\lambda$ and $\nu$ for the three 
different pion beam energies $E_\pi$ as indicated in Fig.~\ref{fig3}. The decrease of $\lambda$ 
with $q_T$ is moderated a bit and the increase of $\nu$ with $q_T$ is strengthened by the Glauber 
effect, such that the measured LT violations $\delta$ are well accommodated. We point out that all 
the functions $H_i(Q^2,y,q_T^2)$ decrease with $q_T$, but $H_\nu(Q^2,y,q_T^2)$ decreases more 
slowly under the Glauber effect, explaining the large enhancement of $\nu$. This feature will be 
illustrated in the geometric picture near the end of this paper. We have confirmed that the 
perturbative results for $\delta$, corresponding to $S=0$, vanish at LO, and coincide with the 
horizontal axes in Fig.~\ref{fig3}. The NLO results for $\delta$, being negative and nearly zero 
with magnitudes smaller than 0.1 in the region $q_T\le 3$ GeV \cite{Chang:2018pvk}, still deviate 
from the data obviously. We remind that the ascent of the curves for $\delta$ in Fig.~\ref{fig3} 
should not extend to the high $q_T$ region, where the collinear factorization holds, and the Glauber 
effect is supposed to diminish. In fact, the LT violation $\delta<0$ with an opposite sign has been 
observed at high $q_T$ of $Z$ boson production in proton-proton collisions \cite{CMS}. As to the 
dependence on the pion beam energy, we find that the results of $\lambda$ ($\nu$) increase (decrease) 
with $E_\pi$ for fixed $q_T$, so those of $\delta$ decrease with $E_\pi$.


\begin{figure}[tb]
\begin{center}
\includegraphics[scale=0.42]{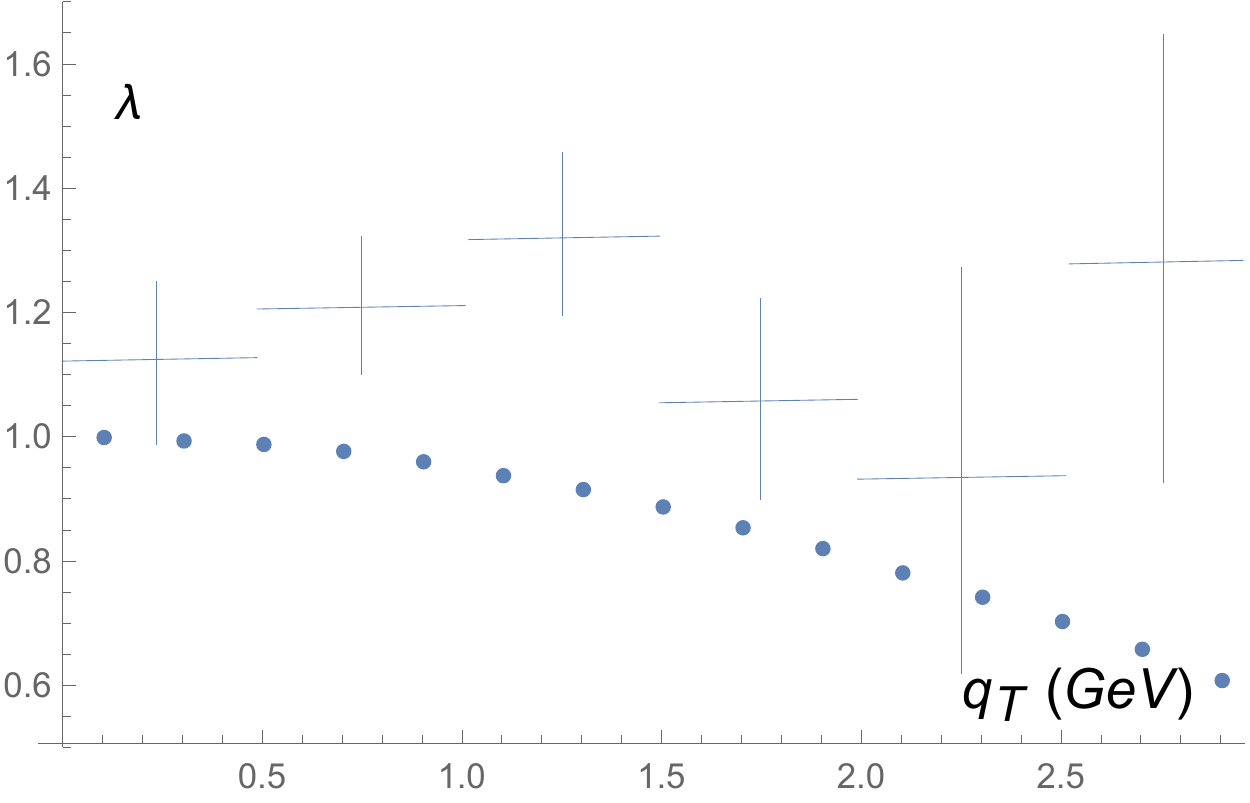}\hspace{0.3 cm}
\includegraphics[scale=0.42]{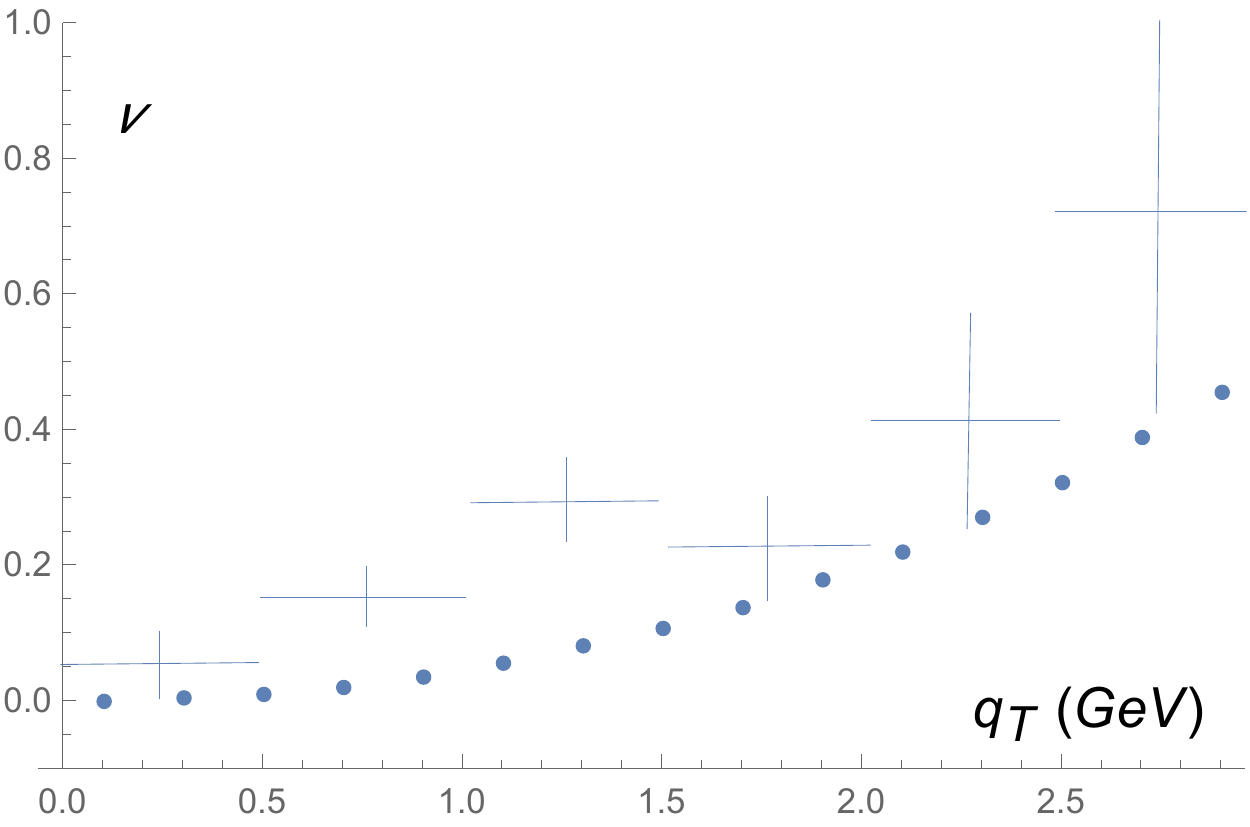}\hspace{0.3 cm}
\includegraphics[scale=0.42]{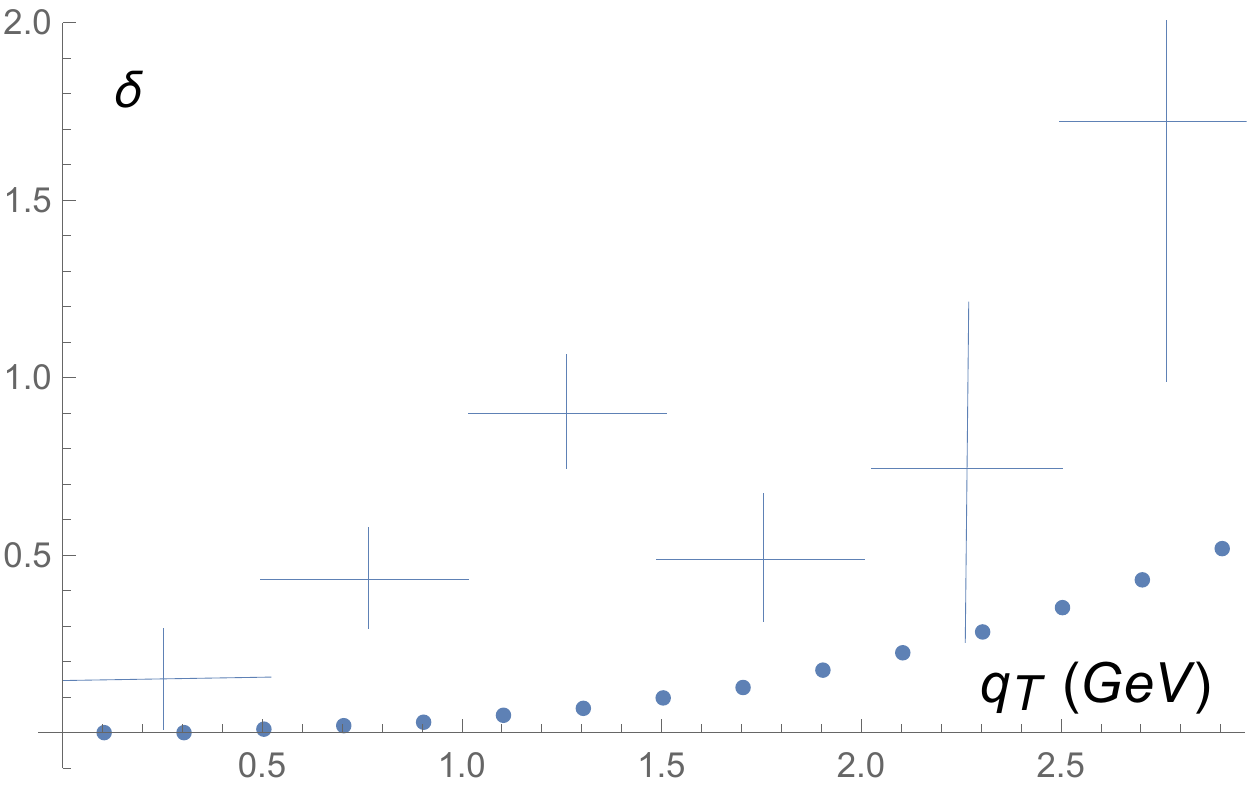}

\caption{Dependencies of $\lambda$, $\nu$, and the LT violation $\delta$ on the
lepton-pair transverse momentum $q_T$, and their comparisons with the E615 data \cite{E615}
for the pion beam energy $E_\pi=252$ GeV.}\label{fig4}
\end{center}
\end{figure}

The kinematic cuts $4.05 \le Q \le 8.55$ GeV, $0.2 \le x_\pi\le 1$ and $0 \le x_F\le 1$ 
were implemented in the E615 experiment with the pion beam energy $E_\pi=252$ GeV \cite{E615}.
We perform the integrations in Eq.~(\ref{ang}) over the ranges of $y$ and $Q$ accordingly,  
\begin{eqnarray}
& &\frac{1}{2}\left(c+\sqrt{c^2+4}\right)\le e^y\le \frac{1}{2}\left(b+\sqrt{b^2-4}\right),
\;\;\;\;{\rm as}\;\;Q^2 \le 0.2^2 s,\nonumber\\
& &1\le e^y\le \frac{1}{2}\left(b+\sqrt{b^2-4}\right),\;\;\;\;{\rm as}\;\;Q^2 > 0.2^2 s,
\nonumber\\
& &c=\frac{0.2^2 s-Q^2}{0.2\sqrt{s(Q^2+q_T^2)}}.\label{xpf4}
\end{eqnarray}
The predicted $q_T$ spectra of the angular coefficients $\lambda$ and $\nu$, and the LT 
violation $\delta$ for the Glauber phase $S=0.8$ under Eq.~(\ref{xpf4}) are shown in 
Fig.~\ref{fig4}, whose behaviors are close to those in Fig.~\ref{fig3}. The discussions 
of the Glauber effect on those $q_T$ spectra also proceed similarly. Our predictions for 
$\nu$ and $\delta$ are slightly lower than the E615 data \cite{E615}, but the consistency 
is still satisfactory, after the sizable experimental errors are considered.
In particular, the deviation from the LT relation, i.e., from the horizontal axis in
the third plot, is roughly accounted for by the Glauber effect.


\begin{figure}[tb]
\begin{center}
\includegraphics[scale=0.42]{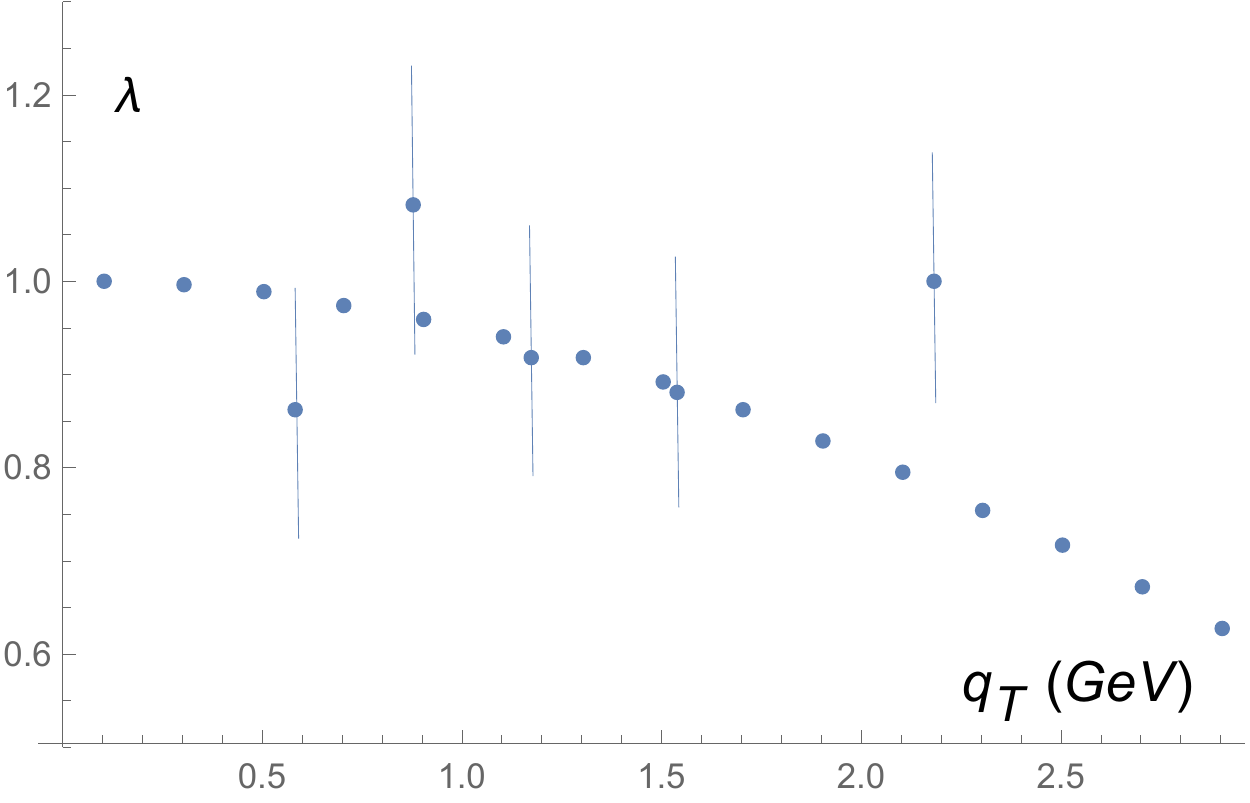}\hspace{0.3 cm}
\includegraphics[scale=0.42]{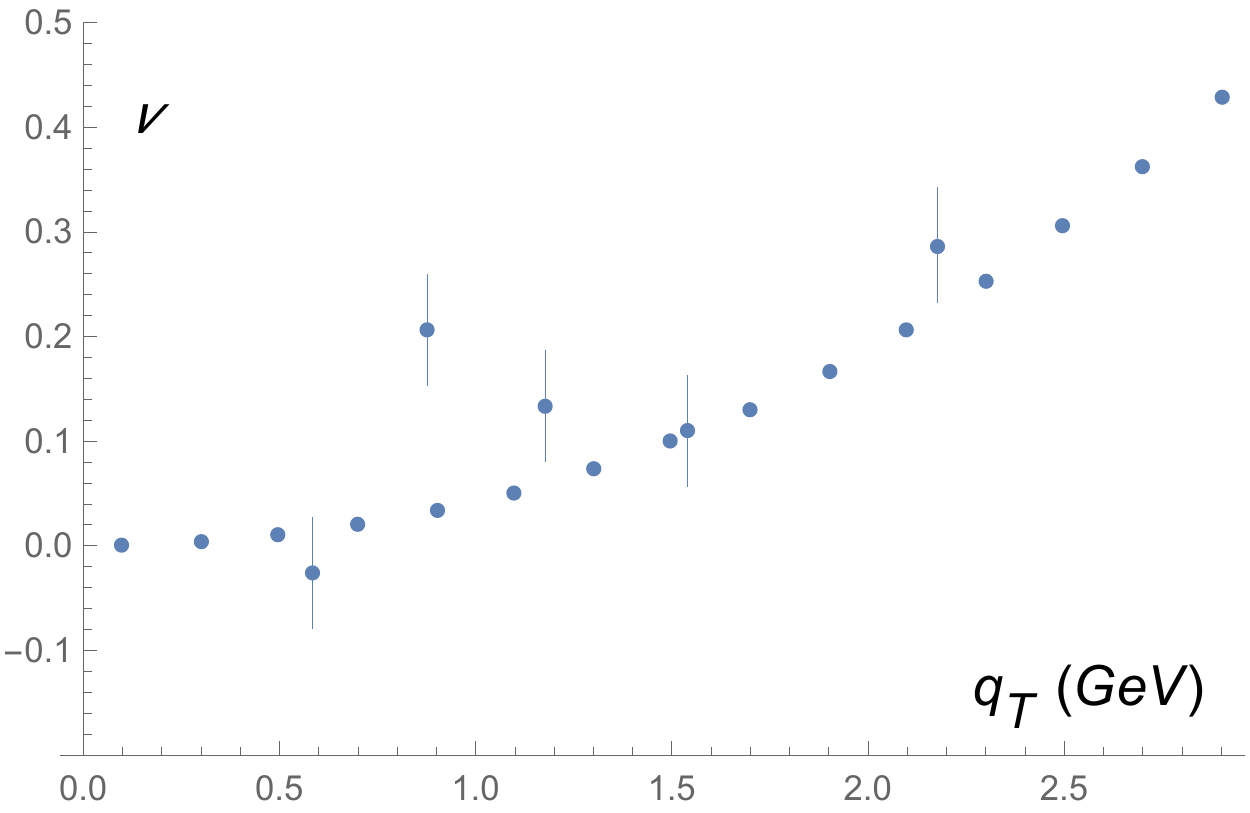}\hspace{0.3 cm}
\includegraphics[scale=0.42]{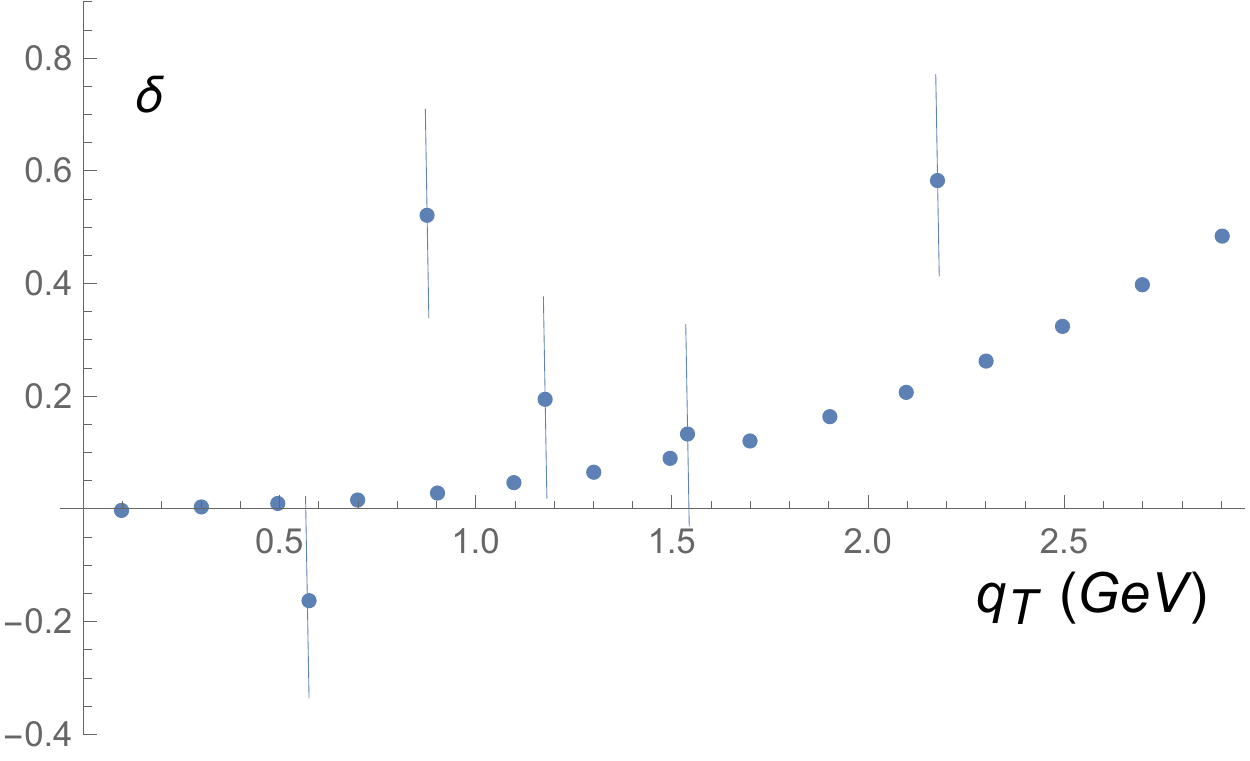}

\caption{Dependencies of $\lambda$, $\nu$, and the LT violation $\delta$ on the
lepton-pair transverse momentum $q_T$, and their comparisons with the COMPASS data \cite{COMPASS}
for the pion beam energy $E_\pi=190$ GeV.}\label{fig5}
\end{center}
\end{figure}

At last, we make predictions for the COMPASS measurements with the pion beam energy 
$E_\pi=190$ GeV. The corresponding cuts 4.3 GeV $\le Q\le$ 8.5 GeV and $x_F\ge -0.1$ 
\cite{COMPASS} lead to the range of $y$,
\begin{eqnarray}
\frac{1}{2}\left(d+\sqrt{d^2+4}\right)\le e^y\le \frac{1}{2}\left(b+\sqrt{b^2-4}\right),
\;\;\;\;d=\frac{-0.1\sqrt{s}}{\sqrt{Q^2+q_T^2}}.
\label{xf3}
\end{eqnarray}
The dependencies of $\lambda$, $\nu$ and $\delta$ on $q_T$, displayed in 
Fig.~\ref{fig5}, are similar to those derived in the previous cases. 
A careful look at Figs.~\ref{fig3}(b) and \ref{fig5} with the approximately 
equal pion beam energies reveals that the values of $\lambda$ ($\nu$) in the former are lower 
(higher) than in the latter. This difference may be attributed to the slightly lower $Q$ 
region that the NA10 measurements have probed. The behaviors of these angular coefficients
in various bins of $Q$, investigated in \cite{Chang:2018pvk}, concur the above tendency. 
Though the preliminary COMPASS data \cite{COMPASS} are not yet precise enough,
the general features remain the same: the decrease (increase) of $\lambda$ 
($\nu$) with $q_T$ is milder (stronger) than expected by the perturbation theory 
\cite{Chang:2018pvk}, and the observed $\delta$, i.e., the deviation from the horizontal
axis in the third plot, is significant. The same Glauber phase also improves the agreement 
between the theoretical results and the COMPASS data for $\lambda$, $\nu$ and $\delta$
simultaneously.

\begin{figure}[tb]
\begin{center}
\includegraphics[scale=0.42]{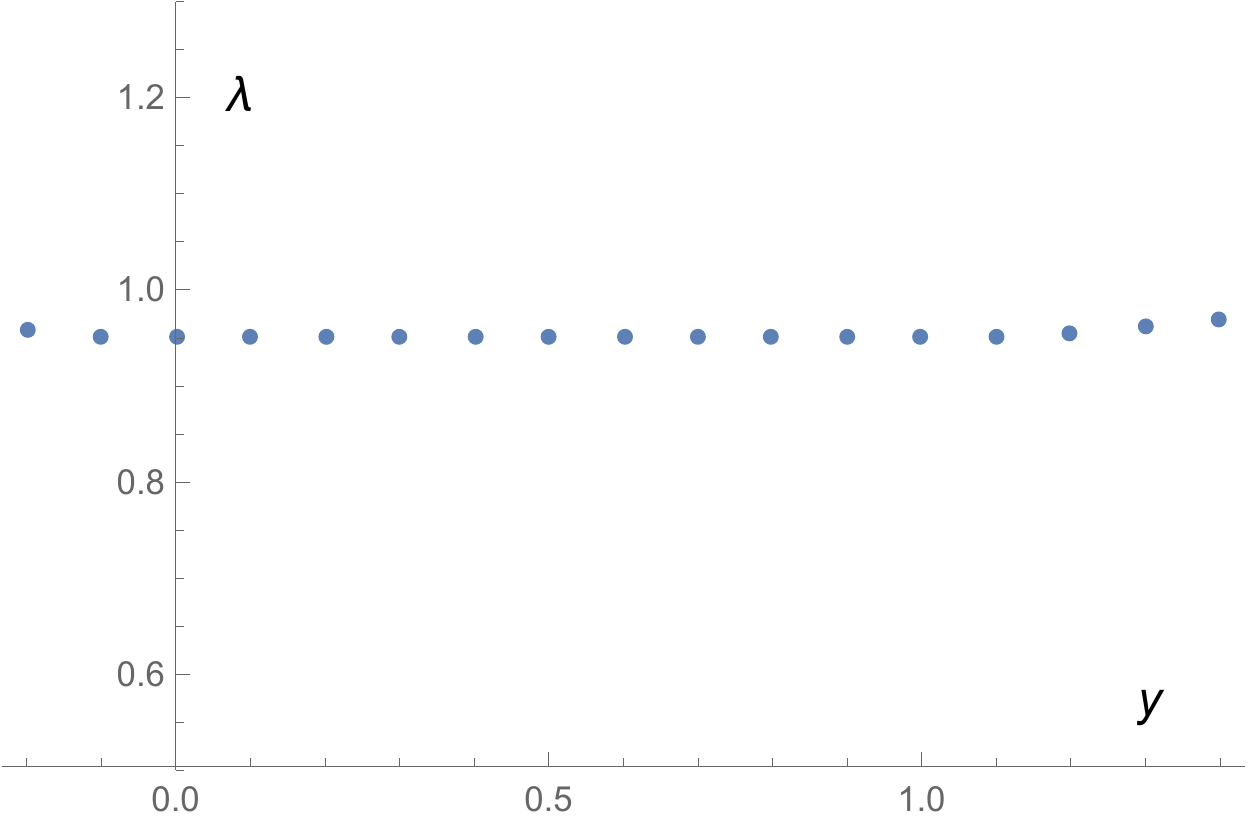}\hspace{0.3 cm}
\includegraphics[scale=0.42]{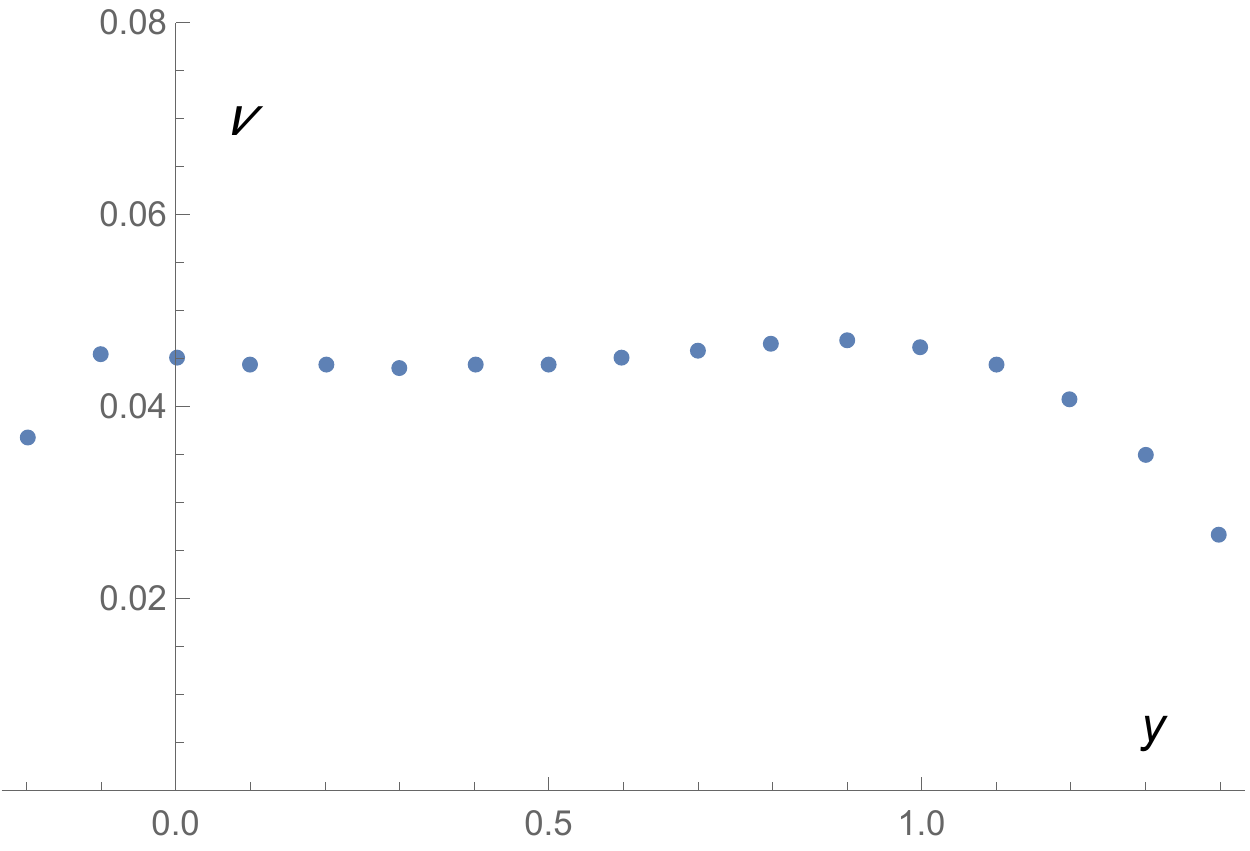}\hspace{0.3 cm}
\includegraphics[scale=0.42]{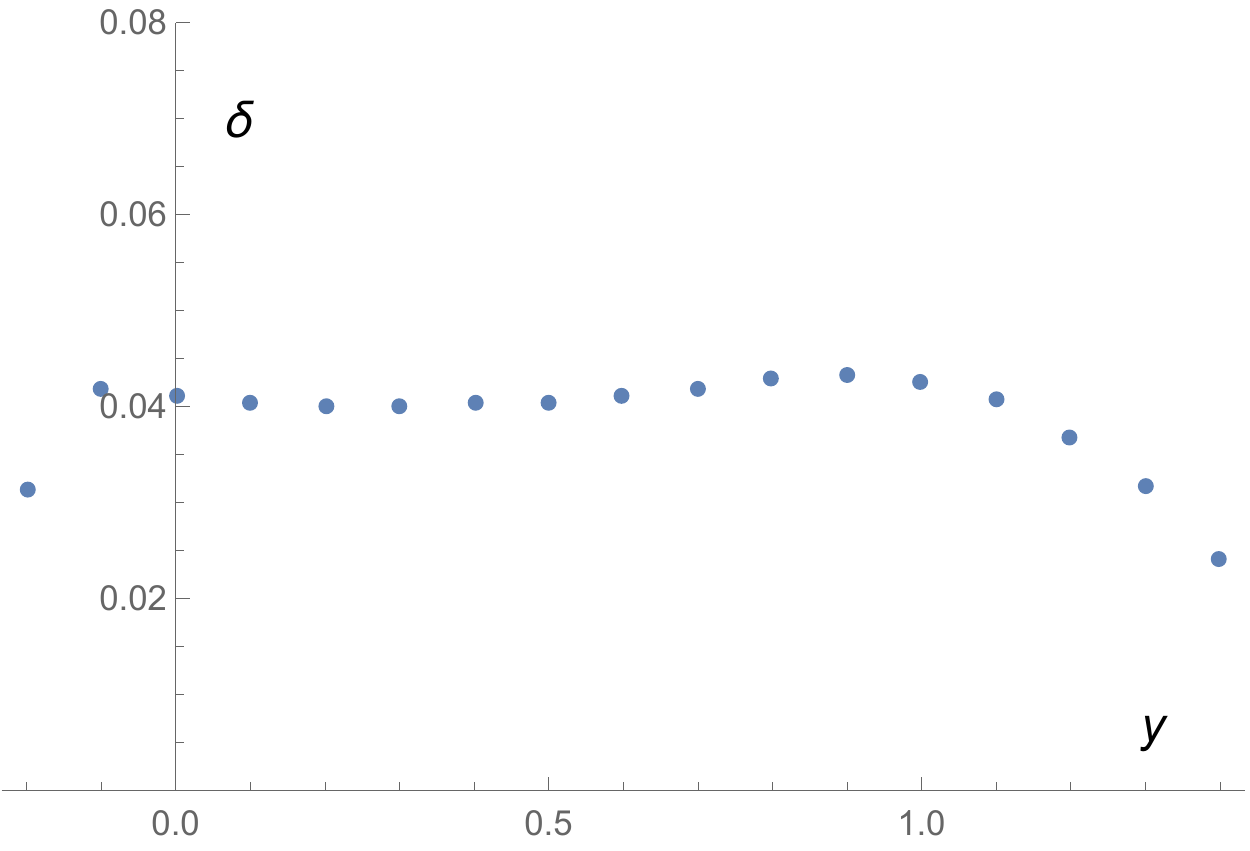}

\caption{Dependencies of $\lambda$, $\nu$, and the LT violation $\delta$ on the
lepton-pair rapidity $y$ for the COMPASS kinematics.}\label{fig6}
\end{center}
\end{figure}

\begin{figure}[tb]
\begin{center}
\includegraphics[scale=0.42]{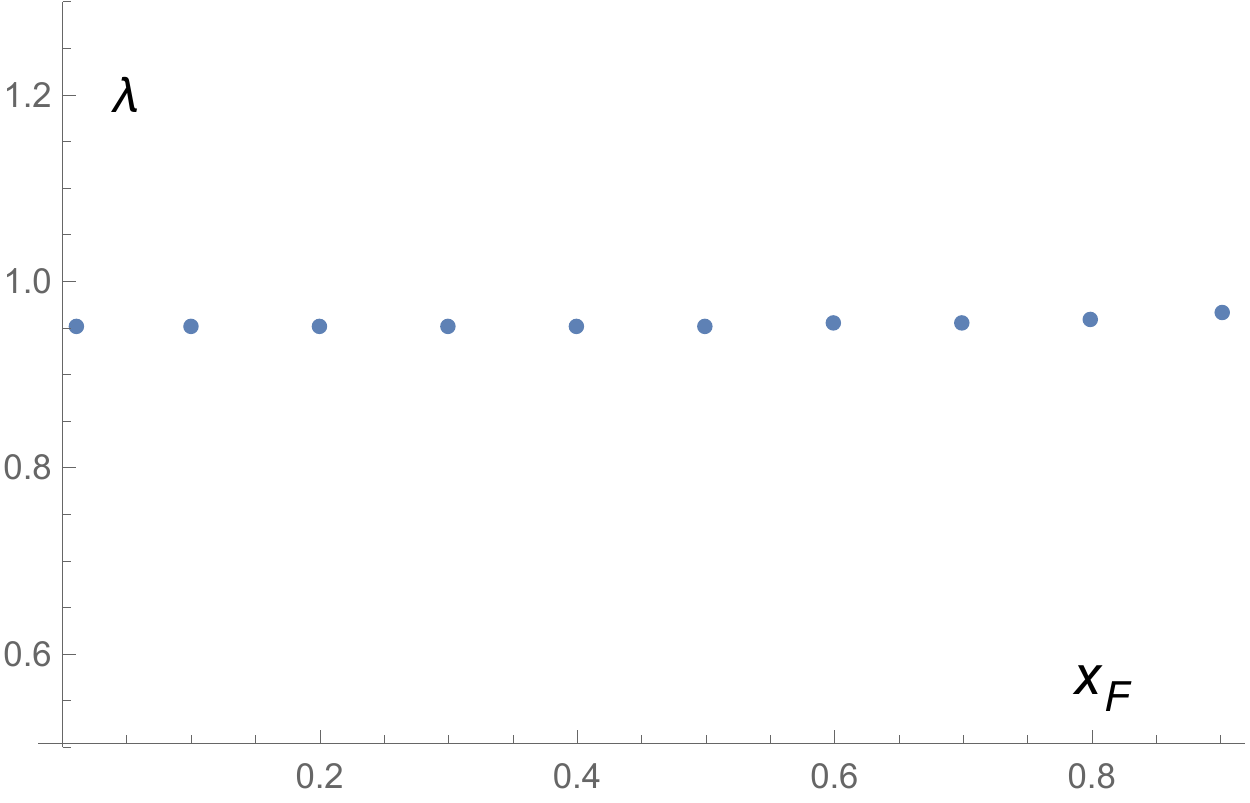}\hspace{0.3 cm}
\includegraphics[scale=0.42]{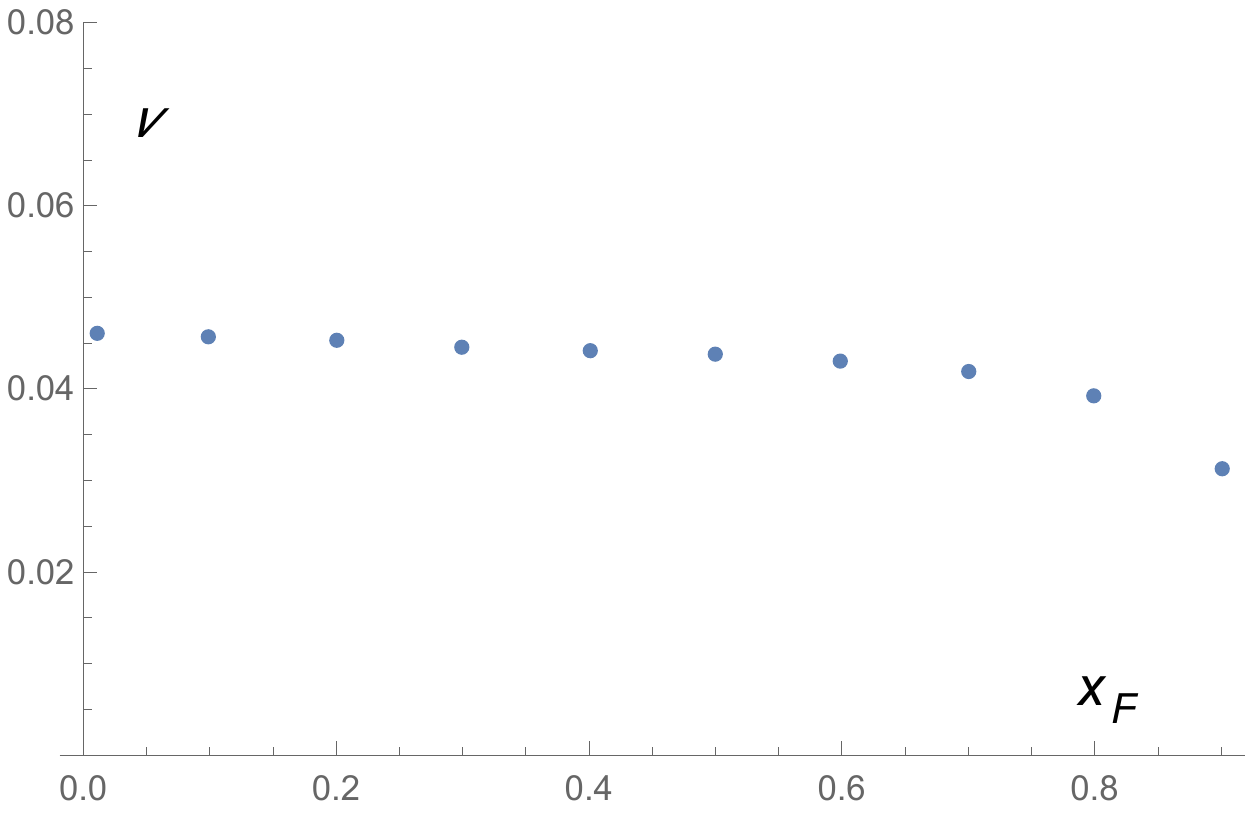}\hspace{0.3 cm}
\includegraphics[scale=0.42]{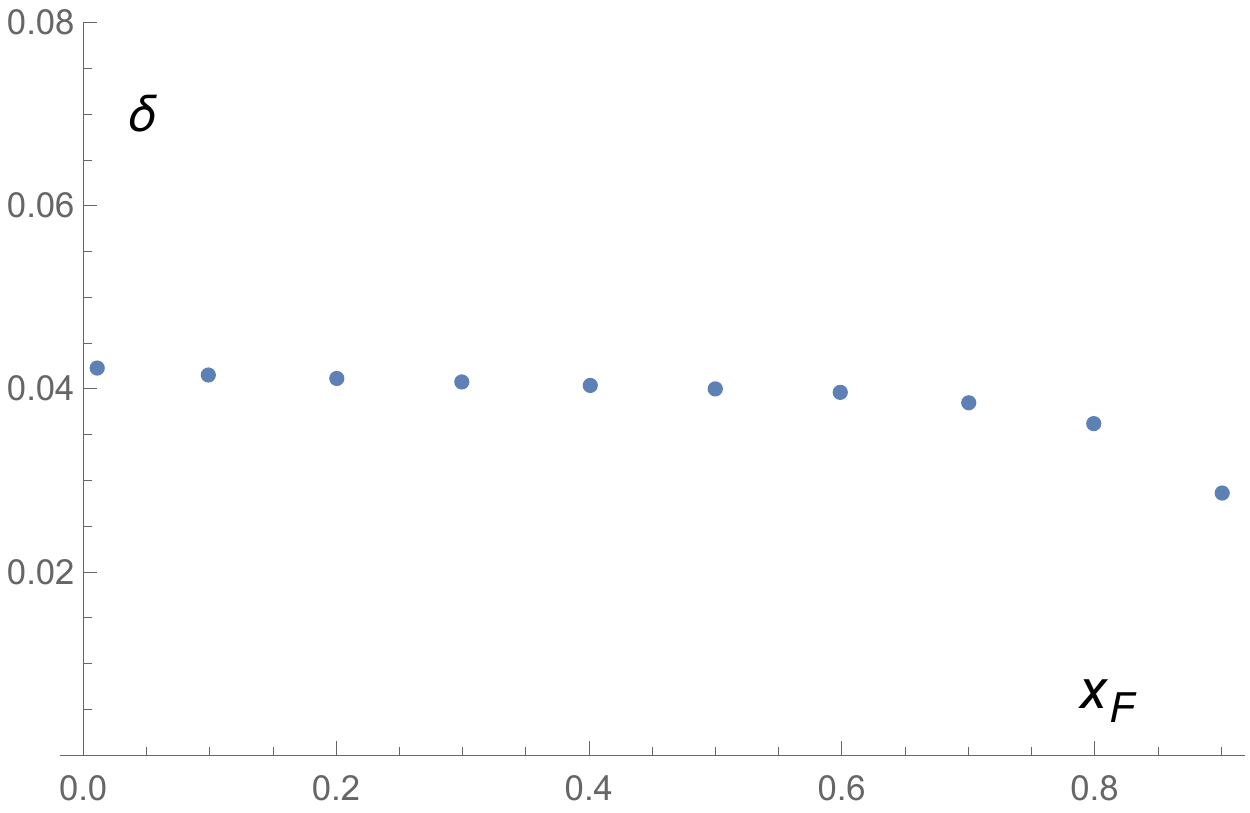}

\caption{Dependencies of $\lambda$, $\nu$, and the LT violation $\delta$ on the
Feynman variable $x_F$ for the COMPASS kinematics.}\label{fig7}
\end{center}
\end{figure}

We also present our predictions for the dependencies of $\lambda$, $\nu$ and $\delta$ 
on the lepton-pair rapidity $y$ in Fig.~\ref{fig6} and on the Feynman variable $x_F$ 
in Fig.~\ref{fig7} for the COMPASS kinematics. In the former
case the definitions of the angular coefficients,
\begin{eqnarray}
\lambda,\mu,\nu=\frac{\int dQ^2dq_T^2 H_{\lambda,\mu,\nu}(Q^2,y,q_T^2)}
{\int dQ^2dq_T^2 H_0(Q^2,y,q_T^2)},
\label{angy}
\end{eqnarray}
are adopted, for which Eq.~(\ref{xf3}) can be converted into the allowed ranges of 
$q_T$ and $Q$ straightforwardly. In the latter case the set of variables $Q$, $y$ and $q_T$ 
has to be changed to the set of $Q$, $y$ and $x_F$ first. The phase space covers the full 
range of $q_T=[0.4, 3.0]$ GeV basically within $y=[-0.1,1.1]$ ($x_F=[0,0.7]$), explaining 
why stable regions exist in $y$ ($x_F$) for $\lambda$, $\nu$ and $\delta$ as shown in 
Fig.~\ref{fig6} (Fig.~\ref{fig7}). Since all the functions $H_i(Q^2,y,q_T^2)$ in Eq.~(\ref{hi}) 
decrease with $q_T$ as stated before, the contributions to the angular coefficients are 
dominated by $q_T<1$ GeV. It is then easy to understand the value of $\lambda$ about 0.95 
and the small values of $\nu$ and $\delta$ around 0.04 in both figures, which are close to 
those for $q_T<1$ GeV in Fig.~\ref{fig5}. The quick descents of $\nu$ and $\delta$ near the 
high ends of $y$ and $x_F$, where the range of $q_T$ shrinks toward small $q_T$, also match 
the results in Fig.~\ref{fig5}.  


It is instructive to examine whether the angular coefficients modified by the
Glauber effect obey the positivity constraints on the rotation-invariant observables
\cite{Ma:2017hfg,Gavrilova:2019jea}, which are defined in terms of the angular coefficients.
Two SO(3) invariants survive in the present case with only virtual photon contributions 
\cite{Gavrilova:2019jea},
\begin{eqnarray}
U_2=\frac{\lambda^2+3\mu^2+3\nu^2/4}{(3+\lambda)^2},\;\;\;\;
T=\frac{(\lambda+3\nu/2)(2\lambda^2+9\mu^2-3\lambda\nu)}{(3+\lambda)^3}.
\end{eqnarray}
To compute the angular coefficients in the above expressions, we integrate $H_i$ in Eq.~(\ref{hi}) over
$Q^2$, $y$ and $q_T^2$, and then take their ratios. Considering the NA10 kinematics for 
the pion beam energy $E_\pi=194$ GeV, and performing the integration over the range of $q_T=[0, 2.5]$
GeV, we find that the values of $\mu$ and $\nu$ are quite small, and $\lambda$ is close to unity. 
They thus lead to $U_2=0.062$ and $T=0.031$, which satisfy the positivity constraints $U_2\le 1/4$ and 
$-1/8\le T\le 3/8$ \cite{Gavrilova:2019jea}, respectively. In addition to the SO(3) invariants,
one can consider the SO(2) invariants \cite{Gavrilova:2019jea}, which are given, in terms of the same
angular coefficients, by
\begin{eqnarray}
& &I_x=\frac{1+\lambda-\nu}{3+\lambda}=0.498,\;\;\;\;
I_y=\frac{1+\lambda+\nu}{3+\lambda}=0.501,\;\;\;\;
I_z=\frac{1-\lambda}{3+\lambda}=0.002,\nonumber\\
& &I_{xx}=\frac{1}{4}(I_x-1)^2-\frac{(\lambda+\nu/2)^2}{(3+\lambda)^2}=0.001,\;\;\;\;
I_{yy}=\frac{1}{4}(I_y-1)^2-\frac{(\lambda-\nu/2)^2+4\mu^2}{(3+\lambda)^2}=0.001,\nonumber\\
& &I_{zz}=\frac{1}{4}(I_z-1)^2-\frac{\nu^2}{(3+\lambda)^2}=0.249.
\end{eqnarray}
The above results also respect the constraints $0\le I_x,I_y,I_z\le 1$ and
$0\le I_{xx},I_{yy},I_{zz}\le 1/4$ \cite{Gavrilova:2019jea}. 

In particular, $I_y$ is identical to the invariant ${\cal F}=(1+\lambda_0)/(3+\lambda_0)$ introduced 
in \cite{Faccioli:2010ej} with the angular coefficient 
\begin{eqnarray}
\lambda_0=\frac{\lambda+3\nu/2}{1-\nu/2},
\end{eqnarray}
in the privileged frame \cite{OVT}. The numerator and the denominator of $\lambda_0$ are expressed as the convolutions
of the PDFs with the hard kernels 
\begin{eqnarray}
\left(\frac{E_1}{E_2}+\frac{E_2}{E_1}\right)\frac{1}{\sin^2\theta_1}
+ \frac{(\cos{S}-1)}{\sin^2\theta_1}\left[\frac{E_1}{E_2}
+\frac{E_2}{E_1}-2\left(\frac{E_1E_2}{k^2}\sin^2\theta_1+1\right)\cos^2\theta_1\right],
\end{eqnarray}
and 
\begin{eqnarray}
\left(\frac{E_1}{E_2}+\frac{E_2}{E_1}\right)\frac{1}{\sin^2\theta_1}
-\frac{(\cos{S}-1)}{\sin^2\theta_1}\left(\frac{E_1-2k}{E_2}
+\frac{E_2-2k}{E_1}-\frac{2E_1E_2}{k^2}\cos^2\theta_1+4\right),
\end{eqnarray}
respectively. As expected, we recover $\lambda_0=1$ as the LT relation holds \cite{OVT}, i.e., as
the Glauber phase $S$ vanishes .

The geometric approach developed in \cite{OVT,PCM,WMP,PBC} has provided a transparent 
illustration of how higher-order corrections in the perturbation theory give rise to 
the LT violation. The argument starts from the lepton pair production in a 
quark-anti-quark annihilation process, which obeys the angular distribution 
$1+\cos^2\theta_0$ with $\theta_0$ being the polar angle of a lepton relative to one of 
the colliding quarks in the rest frame of the lepton pair \cite{FLS}. Only when an 
on-shell quark and an on-shell anti-quark annihilate, does the produced lepton pairs 
obey this simple distribution. Therefore, the geometric picture applies better to the 
region with $q_T$ being much lower than other hard scales like $Q$, in which the colliding 
quarks stay on-shell approximately after radiating collinear gluons. The angle $\theta_0$ 
is then related to $\theta$ and $\phi$ in the CS frame through \cite{PCM,WMP}
\begin{eqnarray}
\cos\theta_0=\cos\theta\cos\theta_1+\sin\theta\sin\theta_1\cos(\phi-\phi_1),\label{t0}
\end{eqnarray}
with $\theta_1$ ($\phi_1$) being the polar (azimuthal) angle of the colliding quark referred to above
in the CS frame. The angle $\theta_1$ has the same meaning as that in Eq.~(\ref{pp}), and
$\phi_1=0$ at LO, i.e., for the $O(\alpha_s)$ diagrams in Fig.~\ref{fig1}.
The angular coefficients $\lambda$ and $\nu$ are expressed in terms of
$\theta_1$ and $\phi_1$ as
\begin{eqnarray}
\lambda=\frac{2-3A_0}{2+A_0},\;\;\;\;\nu=\frac{2A_2}{2+A_0},\label{ln}
\end{eqnarray}
with the functions $A_0=\langle \sin^2\theta_1\rangle$ and  
$A_2=\langle \sin^2\theta_1\cos(2\phi_1)\rangle$,
where the averages are performed over an event sample, i.e., over the corresponding 
differential cross section. 

As stated in \cite{PCM}, the dependence on the azimuthal angle
$\phi_1$ of the quark plane is caused by transverse momenta of radiative gluons, which can 
be achieved at $O(\alpha_s^2)$. Certainly, the values of $\theta_1$ at $O(\alpha_s)$ and at
$O(\alpha_s^2)$ may differ too, but this difference does not affect the reasoning below. 
The inequality $A_2\le A_0$ due to $\cos(2\phi_1)\le 1$ 
then breaks the LT relation, yielding a negative violation $\delta$. The predicted negative 
$\delta$ \cite{Chang:2018pvk}, contrary to the experimental indication of the pion-induced 
Drell-Yan processes at low $q_T$, hints that the LT violation might originate from a 
nonperturbative mechanism. The BM function, as a TMD PDF, invokes the correlation between 
the spin of the colliding quark and its transverse momentum, which modifies the perturbative 
results of $\nu$, but not those of $\lambda$. Hence, it represents an additional contribution 
to the geometric picture, in which the colliding quarks are unpolarized. This is the reason 
why the BM mechanism can stimulate a positive $\delta$ at $O(\alpha_s^0)$ by breaking the 
azimuthal symmetry of the lepton pair distribution, that fits the data of the pion-induced 
Drell-Yan processes. Note that the sign of $\delta$ is not a prediction of the BM proposal,
but a fit from the data.

The Glauber gluon effect on the LO results, different from the above known contributions, 
can also be elaborated in terms of the geometric picture. The necessary rung gluon emission 
on the pion side in Fig.~\ref{fig1}(a), being mainly collinear, tends to decrease the 
anti-quark energy and to lower the lepton-pair invariant mass $Q$. The azimuthal angle 
$\phi_1$ of the quark plane remains tiny under the collinear gluon emission. For a given 
$q_T$, it implies that the mechanism tends to enlarge $\theta_1$, and thus decreases the 
coefficient $\lambda$ and increases $\nu$ in Eq.~(\ref{ln}). A Glauber gluon then injects a 
transverse momentum into the colliding quarks, rendering them off-shell and space-like. The 
produced lepton pairs will follow a modified angular distribution $\epsilon+\cos^2\theta_0$ 
with the parameter $\epsilon < 1$ being attributed to the space-like virtuality of the quarks. 
This modified distribution can be derived trivially by computing the differential cross 
section for the annihilation $\bar q+q \to \ell^- +\ell^+$ with off-shell initial quarks. 
The insertion of Eq.~(\ref{t0}) leads to a smaller denominator $2\epsilon+A_0$ in 
Eq.~(\ref{ln}), such that the net effect makes a minor impact on $\lambda$, but a
strong enhancement of $\nu$. The above simple reasoning elucidates the results in 
Fig.~\ref{fig2}, and the positive deviations $\delta$ derived in our analysis.
We remark that both nontrivial $\epsilon$ and $\phi_1$ can be induced at $O(\alpha_s^2)$
for high $q_T$ in the geometric approach, and this complicated case deserves a thorough discussion.

In this paper we have demonstrated that the Glauber gluon effect, having been employed to resolve 
the several puzzles in the heavy quark decays, can explain the violation of the LT relation in the 
pion-induced Drell-Yan processes. The Glauber phase $S\sim 0.8$ is the only free parameter 
in our formalism, which was fixed by the NA10 data for the angular coefficient $\nu$ with a higher 
precision. This phase was then used to predict the coefficient $\lambda$ and the LT violation
$2\nu+\lambda-1$, which were shown to accommodate all the data from the NA10, E615 and COMPASS 
experiments. Compared to the previous study, we have adopted the realistic PDFs for a proton
from the CT18 and for a pion from the xFitter, included the QCD evolutions of the strong 
coupling and the PDFs, and integrated the differential cross section over the kinematic 
region considered in the above measurements. We have argued that the Glauber effect may be 
significant in pion-induced processes due to the unique role of a pion as a NG boson and a $q\bar q$ 
bound state, and illustrated it in the geometric picture. The distinctions from the perturbative and 
BM mechanisms have been stressed, and measuring the lepton pair distribution in the proton-anti-proton 
Drell-Yan process at low $q_T$ can discriminate the different resolutions for the LT violation. 
Precise data of the coefficient $\lambda$ can also serve the purpose, for which the perturbative and 
BM results stay below those from the Glauber effect. It is mentioned that the angular distribution 
of the lepton pairs in the proton-anti-proton Drell-Yan process produced at the $Z$ pole by the 
CDF \cite{CDF} satisfies the LT relation in the lowest bin of $q_T=0$-10 GeV. It will be an important 
measurement \cite{Frankfurt:2018msx} for exploring the internal structures of hadrons and for 
understanding the correlation of colliding partons in Drell-Yan processes. If the Glauber effect 
associated with a pion turns out to be crucial, it should be included in the extraction of the TMD pion 
PDF from the data of pion induced Drell-Yan processes.

\acknowledgments{
We thank W.C. Chang, T.J. Hou, Y.S. Lian, and J.C. Peng for useful discussions. This work was supported 
by the National Science Council of R.O.C. under the Grant No.
NSC-101-2112-M-001-006-MY3.}

\appendix

\section{Average Glauber Phase}

In this Appendix we explain how to take the average of the Glauber phase $S$ for Figs.~\ref{fig1}(a) 
and \ref{fig1}(b), and how this operation simplifies the corresponding factorization formulas. 
We quote Eq.~(29) of Ref.~\cite{CL09} for the factorization of
one Glauber gluon exchange on the left-hand side of the final-state cut in Fig.~\ref{fig1}(a),
\begin{eqnarray}
T_{L}^{(1)}\approx-i\frac{g^2}{(2\pi)^2}\int
\frac{d^2l_T}{l_T^2+m_g^2}H({\bf
p}_{1T}-{\bf l}_T-{\bf q}_{T},{\bf p}_{1T}-{\bf q}_{T})
\Phi_\pi({\bf p}_{1T}-{\bf l}_T,{\bf p}_{1T})\cdots,\label{ts1}
\end{eqnarray}
where the kinematic variables have been modified to coincide with those in this work, 
and $\cdots$ represents other factors not explicitly shown. The first argument of the hard 
kernel $H$ (the TMD pion PDF $\Phi_\pi$) denotes the transverse momentum of the virtual (valence) 
quark on the left-hand side of the final-state cut, and the second arguments denote those 
on the right-hand side of the cut. Equation~(\ref{ts1}) indicates that the transverse 
momentum $l_T$ of the Glauber gluon flows through the hard kernel $H$, and that an infrared 
divergence is generated from the region $l_T\to 0$ to be regularized by a gluon mass $m_g$. 
The intrinsic transverse momentum $p_{1T}$ in the pion is small, and lower than 1 GeV
typically. Since the lepton-pair transverse momentum $q_{T}$ is about few GeV, at which 
the LT violation is significant, the integrals over $l_T$ and $p_{1T}$ in the dominant Glauber 
region depend on $q_T$ weakly.

We then apply the approximation $H({\bf p}_{1T}-{\bf l}_T-{\bf q}_{T},{\bf p}_{1T}-{\bf q}_{T})
\approx H({\bf q}_{T})$ by neglecting the smaller $l_T$ and $p_{1T}$, 
and Fourier transform Eq.~(\ref{ts1}) into the impact-parameter space,
\begin{eqnarray}
T_{L}^{(1)}\approx\int d^2b_l d^2b_r\Phi_\pi({\bf b}_l,{\bf b}_r)[-iS({\bf
b}_l)]H({\bf q}_{T})e^{i{\bf p}_{1T}\cdot({\bf b}_l-{\bf b}_r)}\cdots,\label{left}
\end{eqnarray}
with the one-loop Glauber factor \cite{CL09}
\begin{eqnarray}
S({\bf b})=\frac{g^2}{(2\pi)^2} \int
\frac{d^2l_T}{l_T^2+m_g^2}e^{-i{\bf l}_T\cdot {\bf
b}}=\frac{g^2}{2\pi} K_0(bm_g),\label{sf}
\end{eqnarray}
$K_0$ being the modified Bessel function. Because the two virtual quarks carry the same transverse 
momentum under the approximation, a single argument ${\bf q}_T$ for $H$ is enough.
The addition of one Glauber gluon to the right-hand side
of the final-state cut, and the extension of the factorization for Glauber gluons to all orders lead to
\begin{eqnarray}
\frac{d\sigma}{dQ^2dy dq_T^2d\Omega}&\approx&\int d^2b_ld^2b_r
\int\frac{d^2k_{3T}}{(2\pi)^2}\Phi_\pi({\bf b}_l,{\bf b}_r)
e^{-iS({\bf b}_l)}H({\bf q}_{T})e^{iS({\bf b}_r)}\nonumber\\
& &\times\Phi_p({\bf b}_l-{\bf b}_r)
e^{i({\bf q}_{T}+{\bf k}_{3T})\cdot({\bf b}_l-{\bf b}_r)},\label{ys}
\end{eqnarray}
which can be deduced straightforwardly by following the steps in Sec.~III of Ref~\cite{CL09}.
Briefly speaking, the $\delta$ function $\delta^2({\bf p}_{1T}+{\bf p}_{2T}-{\bf q}_{T}-{\bf k}_{3T})$ 
for the momentum conservation is integrated over $p_{1T}$, such that 
$e^{i{\bf p}_{1T}\cdot({\bf b}_l-{\bf b}_r)}$ in Eq.~(\ref{ts1})
produces two Fourier factors $e^{-i{\bf p}_{2T}\cdot({\bf b}_l-{\bf b}_r)}$ and
$e^{i({\bf q}_{T}+{\bf k}_{3T})\cdot({\bf b}_l-{\bf b}_r)}$. The former
brings the TMD proton PDF $\Phi_p$ into the impact-parameter space after the integration over $p_{2T}$, 
giving $\Phi_p({\bf b}_l-{\bf b}_r)$. The latter has been kept in Eq.~(\ref{ys}).

The Fourier factor $e^{i({\bf q}_{T}+{\bf k}_{3T})\cdot({\bf b}_l-{\bf b}_r)}$
does not vary much with ${\bf b}_l-{\bf b}_r$ in the region governed by the TMD proton PDF, for
${\bf q}_{T}+{\bf k}_{3T}$ is as small as the intrinsic transverse momenta. This insensitivity 
allows the introduction of a constant Glauber factor $e^{iS}\equiv\langle e^{i[S({\bf b}_r)-S({\bf b}_l)]}\rangle$, 
which is computed as an average over the impact parameters $b_l$ and $b_r$. Noticing that only
the real part $\cos S$ contributes to the differential cross section, we rewrite Eq.~(\ref{ys}) as
\begin{eqnarray}
\frac{d\sigma}{dQ^2dy dq_T^2d\Omega}\approx \cos S\int d^2b_ld^2b_r
\int\frac{d^2k_{3T}}{(2\pi)^2}\Phi_\pi({\bf b}_l,{\bf b}_r)
H({\bf q}_{T})\Phi_p({\bf b}_l-{\bf b}_r)e^{i({\bf q}_{T}+{\bf k}_{3T})\cdot({\bf b}_l-{\bf b}_r)}.\label{ys1}
\end{eqnarray}
The integration of $e^{i{\bf k}_{3T}\cdot({\bf b}_l-{\bf b}_r)}$ over $k_{3T}$ yields 
$(2\pi)^2\delta^2({\bf b}_l-{\bf b}_r)$, which is then integrated over $b_l$ to arrive at
$e^{i{\bf q}_{T}\cdot({\bf b}_l-{\bf b}_r)}=1$ and 
$\Phi_p({\bf b}_l-{\bf b}_r=0)=\phi_p$, i.e., the proton PDF appearing in Eq.~(\ref{4}). At last, Eq.~(40) 
in Ref.~\cite{CL09}, i.e., $\int d^2b_r\Phi_\pi({\bf b}_r,{\bf b}_r)=\phi_\pi$, 
which relates a two-parameter PDF to the corresponding standard PDF, is implemented, and Eq.~(\ref{ys1}) 
reduces to the factorization formula in Eq.~(\ref{4}).

For the addition of Glauber gluons to Fig.~\ref{fig1}(b), we simply replace $H({\bf q}_T)$ in Eq.~(\ref{left})
by $H({\bf k}_{3T})$, and Eq.~(\ref{ys1}) becomes 
\begin{eqnarray}
\frac{d\sigma}{dQ^2dy dq_T^2d\Omega}\approx \cos S\int d^2b_ld^2b_r
\int\frac{d^2k_{3T}}{(2\pi)^2}\Phi_\pi({\bf b}_l,{\bf b}_r)
H({\bf k}_{3T})\Phi_p({\bf b}_l-{\bf b}_r)e^{i({\bf q}_{T}+{\bf k}_{3T})\cdot({\bf b}_l-{\bf b}_r)}.\label{ys2}
\end{eqnarray}
The integration over $k_{3T}$ then transforms $H({\bf k}_{3T})$ into $H({\bf b}_l-{\bf b}_r)$ in the impact-parameter
space. The large scale in the hard kernel, being of order of $Q$, requires ${\bf b}_l\approx{\bf b}_r$,
such that $\Phi_\pi({\bf b}_l,{\bf b}_r)\approx \Phi_\pi({\bf b}_r,{\bf b}_r)$ and 
$\Phi_p({\bf b}_l-{\bf b}_r\approx 0)\approx \phi_p$. The variable change ${\bf b}'_l={\bf b}_l-{\bf b}_r$
and the integration of $H({\bf b}'_l)e^{i{\bf q}_{T}\cdot{\bf b}'_l}$ over $b'_l$ return $H(q_T)$. 
At last, the integration of $\Phi_\pi({\bf b}_r,{\bf b}_r)$ over $b_r$ gives $\phi_\pi$, and we are again led
to the factorization formula in Eq.~(\ref{4}).

\end{document}